\documentclass[lettersize,hidelinks,journal]{IEEEtran}

\usepackage{amsmath,amssymb,amsfonts}
\usepackage{array}
\usepackage{subfigure}
\usepackage{textcomp}
\usepackage{stfloats}
\usepackage{url}
\usepackage{verbatim}
\usepackage{graphicx}
\usepackage{cite}
\usepackage{xcolor}
\usepackage{orcidlink}
\usepackage[acronym,shortcuts]{glossaries}
\usepackage{bm}
\usepackage{relsize}
\usepackage{soul}
\usepackage{algorithm, algpseudocode}
\usepackage{algcompatible}
\usepackage{mathtools}
\usepackage{bm}
\usepackage{lipsum}
\usepackage{mathtools}
\usepackage{lipsum}
\usepackage[hang,flushmargin]{footmisc}
\usepackage{tabularx}

\def\BibTeX{{\rm B\kern-.05em{\sc i\kern-.025em b}\kern-.08em
T\kern-.1667em\lower.7ex\hbox{E}\kern-.125emX}}

\newcommand{\trans}[0]{^{\mathsf{T}}}

\newcommand{\herm}[0]{^{\mathsf{H}}}

\newcommand{\Real}[1]{\Re\{{#1}\}}
\newcommand{\Imag}[1]{\Im\{{#1}\}}

\newacronym{PCA}{PCA}{principal component analysis}
\newacronym{MDSM}{MDSM}{multi-domain sparse modulation}
\newacronym{P2P}{P2P}{point-to-point}
\newacronym{OTAC}{AirComp}{over-the-air computing}
\newacronym{TX}{TX}{transmitter}
\newacronym{RX}{RX}{receiver}
\newacronym{IoT}{IoT}{Internet of Things}
\newacronym{AI/ML}{AI/ML}{artifitial intelligence/machine learning}
\newacronym{SDR}{SDR}{semi-definite relaxation}
\newacronym{EVD}{EVD}{eigenvalue decomposition}
\newacronym{GR}{GR}{Gaussian randomization}
\newacronym{SCA}{SCA}{successive convex approximation}
\newacronym{BnB}{BnB}{branch and bound}
\newacronym{QT}{QT}{quadratic transform}
\newacronym{RQ}{RQ}{Rayleigh quotient}
\newacronym{SOCP}{SOCP}{second-order cone programming}
\newacronym{CDF}{CDF}{cumulative distribution function}
\newacronym{UF}{UF}{uniform-forcing}
\newacronym{AP}{AP}{access point}
\newacronym{RSDR}{R-SDR}{regularized semi-definite relaxation}
\newacronym{R-SDR}{R-SDR}{regularized SDR}
\newacronym{flops}{flops}{floating point operations}
\newacronym{ED}{ED}{edge device}
\newacronym{SINR}{SINR}{signal to interference-plus-noise ratio}
\newacronym{SIC}{SIC}{successive interference cancellation}
\newacronym{CSI}{CSI}{channel state information}
\newacronym{LoS}{LoS}{line-of-sight}
\newacronym{NLoS}{NLoS}{non-LoS}
\newacronym{RPE}{RPE}{radar parameter estimation}
\newacronym{OTFS}{OTFS}{orthogonal time frequency space}
\newacronym{AFDM}{AFDM}{affine frequency division multiplexing}
\newacronym{CRLB}{CRLB}{Cram{\`e}r-Rao lower bound}
\newacronym{BCRLB}{BCRLB}{Bayesian Cram{\`e}r-Rao lower bound}
\newacronym{BBI}{BBI}{Bayesian bilinear inference}
\newacronym{AoA}{AoA}{angle-of-arrival}
\newacronym{SNR}{SNR}{signal-to-noise ratio}
\newacronym{ML}{ML}{maximum likelihood}
\newacronym{MIMO}{MIMO}{multiple-input multiple-output}
\newacronym{MISO}{MISO}{multiple-input single-output}
\newacronym{SIMO}{SIMO}{single-input multiple-output}
\newacronym{SISO}{SISO}{single-input single-output}
\newacronym{MUSIC}{MUSIC}{multiple signal classification}
\newacronym{MU}{MU}{multi-user}
\newacronym{ROOT-MUSIC}{ROOT-MUSIC}{ROOT multiple signal classification}
\newacronym{JCAS}{JCAS}{joint communication and sensing}
\newacronym{JCR}{JCR}{joint communications and radar}
\newacronym{ISAC}{ISAC}{integrated sensing and communications}
\newacronym{3D}{3D}{three-dimensional}
\newacronym{2D}{2D}{two-dimensional}
\newacronym{1D}{1D}{one-dimensional}
\newacronym{BF}{BF}{beamforming}
\newacronym{ROI}{ROI}{region of interest}
\newacronym{mmWave}{mmWave}{millimeter-wave}
\newacronym{MF}{MF}{matched-filter}
\newacronym{DD}{DD}{delay-Doppler}
\newacronym{SotA}{SotA}{state-of-the-art}
\newacronym{ULA}{ULA}{uniform linear array}
\newacronym{QAM}{QAM}{quadrature amplitude modulation}
\newacronym{ISFFT}{ISFFT}{inverse symplectic finite Fourier transform}
\newacronym{SFFT}{SFFT}{symplectic finite Fourier transform}
\newacronym{ISI}{ISI}{inter-symbol interference}
\newacronym{AWGN}{AWGN}{additive white Gaussian noise}
\newacronym{MSE}{MSE}{mean-squared-error}
\newacronym{LMMSE}{LMMSE}{linear minimum mean square error}
\newacronym{RMSE}{RMSE}{root mean square error}
\newacronym{ESPRIT}{ESPRIT}{estimation of signal parameters via rotational invariant techniques}
\newacronym{OFDM}{OFDM}{orthogonal frequency division multiplexing}
\newacronym{OCDM}{OCDM}{orthogonal chirp division multiplexing}
\newacronym{BS}{BS}{base station}
\newacronym{UE}{UE}{user equipment}
\newacronym{JCEDD}{JCEDD}{joint channel estimation and data detection}
\newacronym{PDA}{PDA}{probabilistic data association}
\newacronym{PMF}{PMF}{probability mass function}
\newacronym{PBiGaBP}{PBiGaBP}{parametric bilinear Gaussian belief propagation}
\newacronym{PBiGAMP}{PBiGAMP}{parametric bilinear generalized approximate message passing}
\newacronym{GaBP}{GaBP}{Gaussian belief propagation}
\newacronym{FT}{FT}{frequency-time}
\newacronym{DFT}{DFT}{discrete Fourier transform}
\newacronym{IDFT}{IDFT}{inverse discrete Fourier transform}
\newacronym{TD}{TD}{time domain}
\newacronym{wlg}{w.l.g.}{without loss of generality}
\newacronym{CP}{CP}{cyclic prefix}
\newacronym{DAF}{DAF}{discrete affine Fourier}
\newacronym{DAFT}{DAFT}{discrete affine Fourier transform}
\newacronym{IDAFT}{IDAFT}{inverse discrete affine Fourier transform}
\newacronym{CPP}{CPP}{\textit{chirp-periodic} prefix}
\newacronym{IDZT}{IDZT}{inverse discrete Zak transform}
\newacronym{DZT}{DZT}{discrete Zak transform}
\newacronym{P/S}{P/S}{parallel-to-serial}
\newacronym{S/P}{S/P}{serial-to-parallel}
\newacronym{SBL}{SBL}{sparse Bayesian learning}
\newacronym{MPA}{MPA}{message passing algorithms}
\newacronym{EM}{EM}{expectation maximization}
\newacronym{sIC}{soft IC}{soft interference cancellation}
\newacronym{soft RG}{soft RG}{soft replica generation}
\newacronym{BG}{BG}{belief generation}
\newacronym{SGA}{SGA}{scalar Gaussian approximation}
\newacronym{CLT}{CLT}{central limit theorem}
\newacronym{PDF}{PDF}{probability density function}
\newacronym{QPSK}{QPSK}{quadrature phase-shift keying}
\newacronym{ICI}{ICI}{inter-carrier interference}
\newacronym{BER}{BER}{bit error rate}
\newacronym{DoF}{DoF}{degrees-of-freedom}
\newacronym{VGA}{VGA}{vector Gaussian approximation}
\newacronym{FD}{FD}{full-duplex}
\newacronym{NMSE}{NMSE}{normalized mean square error}
\newacronym{KL}{KL}{Kullback-Leibler}
\newacronym{LASSO}{LASSO}{least absolute shrinkage and selection operator}
\newacronym{FP}{FP}{fractional programming}
\newacronym{CC}{CC}{communication-centric}
\newacronym{RC}{RC}{raised-cosine}
\newacronym{RRC}{RRC}{root raised-cosine}
\newacronym{6G}{6G}{sixth-generation}
\newacronym{V2X}{V2X}{vehicle-to-everything}
\newacronym{LEO}{LEO}{low-earth orbit}
\newacronym{I/O}{I/O}{input-output}
\newacronym{CE}{CE}{channel estimation}
\newacronym{ICC}{ICC}{integrated communication and computing}
\newacronym{ISCC}{ISCC}{integrated sensing, communications and computing}
\newacronym{PAM}{PAM}{pulse amplitude modulation}
\newacronym{iid}{i.i.d.}{independent and identically distributed}
\newacronym{MEC}{MEC}{mobile edge computing}
\newacronym{REMS}{REMS}{reconfigurable electromagnetic structure}
\newacronym{RIS}{RIS}{reconfigurable intelligent surface}
\newacronym{MMSE}{MMSE}{minimum mean square error}

\begin{document}

\title{A Flexible Design Framework for Integrated Communication and Computing Receivers}

\author{Kuranage Roche Rayan Ranasinghe\textsuperscript{\orcidlink{0000-0002-6834-8877}}, \IEEEmembership{Graduate Student Member,~IEEE,}
Kengo Ando\textsuperscript{\orcidlink{0000-0003-0905-2109}}, \IEEEmembership{Member,~IEEE,}\\
Hyeon Seok Rou\textsuperscript{\orcidlink{https://orcid.org/0000-0003-3483-7629}}, \IEEEmembership{Member,~IEEE,} Giuseppe Thadeu Freitas de Abreu\textsuperscript{\orcidlink{0000-0002-5018-8174}}, \IEEEmembership{Senior Member,~IEEE,} \\
Takumi Takahashi\textsuperscript{\orcidlink{0000-0002-5141-6247}}, \IEEEmembership{Member,~IEEE,} Marco Di Renzo\textsuperscript{\orcidlink{0000-0003-0772-8793}}, \IEEEmembership{Fellow,~IEEE,} \\
and David~Gonz{\'a}lez~G.\textsuperscript{\orcidlink{0000-0003-2090-8481}}, \IEEEmembership{Senior Member,~IEEE}
\thanks{~~K.~R.~R.~Ranasinghe, K.~Ando, H.~S.~Rou and G.~T.~F.~de~Abreu are with the School of Computer Science and Engineering, Constructor University, Campus Ring 1, 28759 Bremen, Germany (emails: [kranasinghe, hrou, gabreu]@constructor.university, k.ando@ieee.org).}
\thanks{~~T.~Takahashi is with the Graduate School of Engineering, Osaka University, Suita 565-0871, Japan (e-mail: takahashi@comm.eng.osaka-u.ac.jp).}
\thanks{~~M.~Di~Renzo is with Universit{\'e} Paris-Saclay, CNRS, CentraleSup{\'e}lec, Laboratoire des Signaux et Syst{\`e}mes, 3 Rue Joliot-Curie, 91192 Gif-sur-Yvette, France and with King's College London, Centre for Telecommunications Research - Department of Engineering, WC2R 2LS London, United Kingdom (e-mails: marco.di-renzo@universite-paris-saclay.fr, marco.di\_renzo@kcl.ac.uk).}
\thanks{~~D.~Gonz{\'a}lez~G. is with the Wireless Communications Technologies Group, Continental AG, Wilhelm-Fay Strasse 30, 65936, Frankfurt am Main, Germany (e-mail: david.gonzalez.g@ieee.org).}
\thanks{~~Parts of this paper have been presented at the International Conference on Computing, Networking and Communication (ICNC), 2025 \cite{RanasingheICNC_comp_2025}.}
\vspace{-2ex}
}

\maketitle

\begin{abstract}
We propose a framework to design \ac{ICC} receivers capable of simultaneously detecting data symbols and performing \ac{OTAC} in a manner that: a) is systematically generalizable to any nomographic function, b) scales to a massive number of \acp{UE} and \acp{ED}, c) supports the computation of multiple independent functions (streams), and d) operates in a multi-access fashion whereby each transmitter can choose to transmit either data symbols, computing signals or both.
For the sake of illustration, we design the proposed multi-stream and multi-access method under an uplink setting, where multiple single-antenna \acp{UE}/\acp{ED} simultaneously transmit data and computing signals to a single multiple-antenna \ac{BS}/\ac{AP}.
Under the communication functionality, the receiver aims to detect all independent communication symbols while treating the computing streams as aggregate interference which it seeks to mitigate; and conversely, under the computing functionality, to minimize the distortion over the computing streams while minimizing their mutual interference as well as the interference due to data symbols.
To that end, the design leverages the \ac{GaBP} framework relying only on element-wise scalar operations coupled with closed-form combiners purpose-built for the \ac{OTAC} operation, which allows for its use in massive settings, as demonstrated by simulation results incorporating up to 200 antennas and 300 \acp{UE}/\acp{ED}. 
The efficacy of the proposed method under different loading conditions is also evaluated, with the performance of the scheme shown to approach fundamental limiting bounds in the under/fully loaded cases.

\end{abstract}

\vspace{-1ex}
\begin{IEEEkeywords}
\ac{ICC}, \ac{GaBP}, Over-the-Air Computing, opportunistic, massive, robust, multi-stream and multi-access.
\end{IEEEkeywords}

\glsresetall

\IEEEpeerreviewmaketitle

\vspace{-2ex}
\section{Introduction}
\label{sec:introduction}

The \ac{6G} of wireless networks \cite{Cheng-XiangCST2023,BrintonCOMAG2025,GonzalezPrelcicVTM2025} is expected to bring about a new era of wireless technologies, where \ac{ISCC} functionalities will enable a wide range of emerging applications such as autonomous driving \cite{GonzalezJOCN2025}, drone swarming \cite{KhaledISPA2024}, digital twins \cite{MasaracchiaCOMST2025}, the \ac{IoT} \cite{ShafiquACCESS20} and more.

This multifunctional perspective of wireless systems has been the subject of intense research in the last few years, although \ac{ISAC} \cite{GaudioTWC2020,Wild_Access_2021,LiuJSAC_ISAC2022, Rou_Asilomar22, Mai_CAMSAP23, Furhling_JISPIN23, Ranasinghe_ICASSP_2024, Bemani_WCL_2024, RanasingheARXIV2024,Gonzalez_ProcIEEE_2024, RanasingheARXIV_blind_2024,Rou_SPM24, Rou_TWC24, Zhang_WC_2024, RexhepiARXIV2024} and \ac{OTAC} \cite{NazerTIT07, Wang_arxiv_2024} were initially investigated somewhat in separate.
More recently, however, the integration of sensing, communication and computing functionalities has attracted increasing attention \cite{DuIoTJ2024, WangMICCIS2024, DongIoTJ2024, WenCOMST2024}, perhaps in recognition to the fact that the computing functionality can also assist with the other functionalities, especially sensing, as discussed in \cite{ZhuIoTJ2019}.
As a consequence, a trend can be observed in recent related literature, to move beyond the earlier focus on theoretical analysis \cite{LiuTWC20, QinWCL21}, transmitter design \cite{QiaoLanWCL2020}, and \ac{BF} algorithms \cite{ChenWCL18,FangSPAWC21, AndoCAMSAP2023, ChenTVT2024, LiWCL2025}, towards the integration aspect of the problem.

As an example, a novel \ac{OTAC} scheme based on digital signals was proposed in \cite{LiuTWC2025} where the authors describe an optimal maximum a posteriori detector to compute aggregated digital symbols.
Yet another example is the work in \cite{HuangTWC2025}, where the computing signals are embedded into an \ac{OTFS} structure, such that a purpose-designed detector for the average \ac{OTFS} receive signals amount to an \ac{OTAC} operation.
Although these contributions do not address integration directly, they facilitate \ac{ICC} in so far as the \ac{OTAC} operation is cast as a detection problem similar to that of symbol estimation in conventional communication systems.
Still, these preceding works do not consider the problem of \ul{simultaneously} detecting communication symbols, and aggregating computing signals towards the computation of nomographic functions, which is the actual core of the \ul{integrated}\footnote{Throughout this work, we consider that \acl{ICC} refers to the simultaneous detection of data and \acl{OTAC} operation, which is distinct from the \ul{co-existence} of such functionalities as found in some \ac{SotA} \ac{ICC} methods \cite{YeSPAWC2024, HeTWC2024}.} communication and computing paradigm.

In view of the above, this article proposes a novel \ac{GaBP}-based \cite{bickson2009gaussian, LiTSP2024, takahashi2018design} receiver design framework for \ac{ICC}, in which both data symbols and computing functions are estimated, in order to yield effective physical layer joint communication and computing functionalities\footnote{The integration of sensing functionality into the \ac{ICC} framework here described -- which would yield an \ac{ISCC} scheme -- can be trivially achieved under the assumption that the communication waveforms can be used for sensing purposes \cite{GaudioTWC2020, Rou_SPM24, Ranasinghe_ICASSP_2024, Bemani_WCL_2024, RanasingheARXIV2024,Gonzalez_ProcIEEE_2024, RanasingheARXIV_blind_2024}. 
Due to space limitations, however, we leave such a variation of the proposed method to be addressed in a follow-up work.}. 

To this end, we first formulate a system model in which any \ac{UE} or \ac{ED} can simultaneously transmit communication and computing signals, both of which are to be detected by the receiving \ac{BS} or \ac{AP}.
We then derive the relevant message passing rules to extract the separate data symbols and computing streams via \ac{GaBP}, which is supported via a closed-form solution to an optimal combiner design problem with \ac{SIC} and \ac{EM}.

Unlike our preliminary work \cite{RanasingheICNC_comp_2025}, where message passing rules were designed to extract all individual elements of the computing stream, here, the \ac{GaBP} receiver is designed to detect, besides data symbols, only the aggregate signals corresponding to each computing stream\footnote{The approach from \cite{RanasingheICNC_comp_2025} will be retained as benchmark method for the purpose of performance assessment.}, in line with traditional \ac{OTAC} approaches \cite{LiuTWC20}.
It can also be understood from the derivations that the \ac{SIC}-enabled combiner functions for any desired arbitrary nomographic function\footnote{We emphasize that this may require the computation of corresponding prior distributions, and their possible representation in terms of Gaussian mixtures, on a case-by-case basis.} -- such as those listed in \cite{perezneira2024waveformscomputingair} -- can be implemented via the subsequent design of the corresponding pre- and post-processing functions.
In addition, as a consequence of the low complexity inherent to the \ac{GaBP} approach, the method can be scaled to massive setups, as demonstrated by simulations shown for systems with up to 200 antennas at the \ac{BS}/\ac{AP} and 300 \acp{UE}/\acp{ED}.
As a result of the strategy, the proposed framework can be applied systematically to simultaneously compute multiple nomographic function streams, concomitant with multiple data streams, resulting in a true, scalable and flexible \ac{ICC} scheme.
%

All in all, our contributions can be summarized as follows:
\setcounter{footnote}{\thefootnote-1}
\begin{itemize}
\item A novel \ac{GaBP}-based flexible receiver framework for \ac{ICC} that can be applied to estimate an arbitrary\footnotemark\, nomographic function output and scaled to massive setups is presented.
\item A novel \ac{GaBP}/\ac{EM}-based \ac{OTAC} combiner design, intrinsic to the framework and free of matrix inversions typical of \ac{SotA} combiners, is derived.
\item The proposed framework is extended to support the simultaneous computing of multiple \ac{OTAC} streams in a free-access fashion whereby each transmitter can transmit data symbols, computing signals or both.
\end{itemize}

The remainder of the article can be summarized as follows.
The system model is described in Section \ref{sec:system_model}, and 
an initial benchmarking scheme is then introduced in Section \ref{sec:benchmarking_scheme}.
Next, the proposed \ac{ICC} procedure is described in Section \ref{sec:proposed_method}.
Finally, Section \ref{sec:multi_stream_ICC} generalizes the proposed method to enable multiple computation streams and the multi-access variation.

\textit{Notation:} Throughout the manuscript, vectors and matrices are represented by lowercase and uppercase boldface letters, respectively;
$\mathbf{I}_M$ denotes an identity matrix of size $M$ and $\mathbf{1}_M$ denotes a column vector composed of $M$ ones; 
the Euclidean norm and the absolute value of a scalar are respectively given by $\|\cdot\|_2$ and $|\cdot|$;
the transpose and hermitian operations follow the conventional form $(\cdot)\trans$ and $(\cdot)\herm$, respectively;
$\Re{\{\cdot\}}$, $\Im{\{\cdot\}}$ and  $\mathrm{min}(\cdot)$ represents the real part, imaginary part and the minimum operator, respectively.
Finally, $\sim \mathcal{CN}(\mu,\sigma^2)$ denotes the complex Gaussian distribution with mean $\mu$ and variance $\sigma^2$, where $\sim$ denotes ``is distributed as''.

\vspace{-0.5ex}
\section{System Model}
\label{sec:system_model}

\begin{figure}[H]
\centering
\includegraphics[width=1\columnwidth]{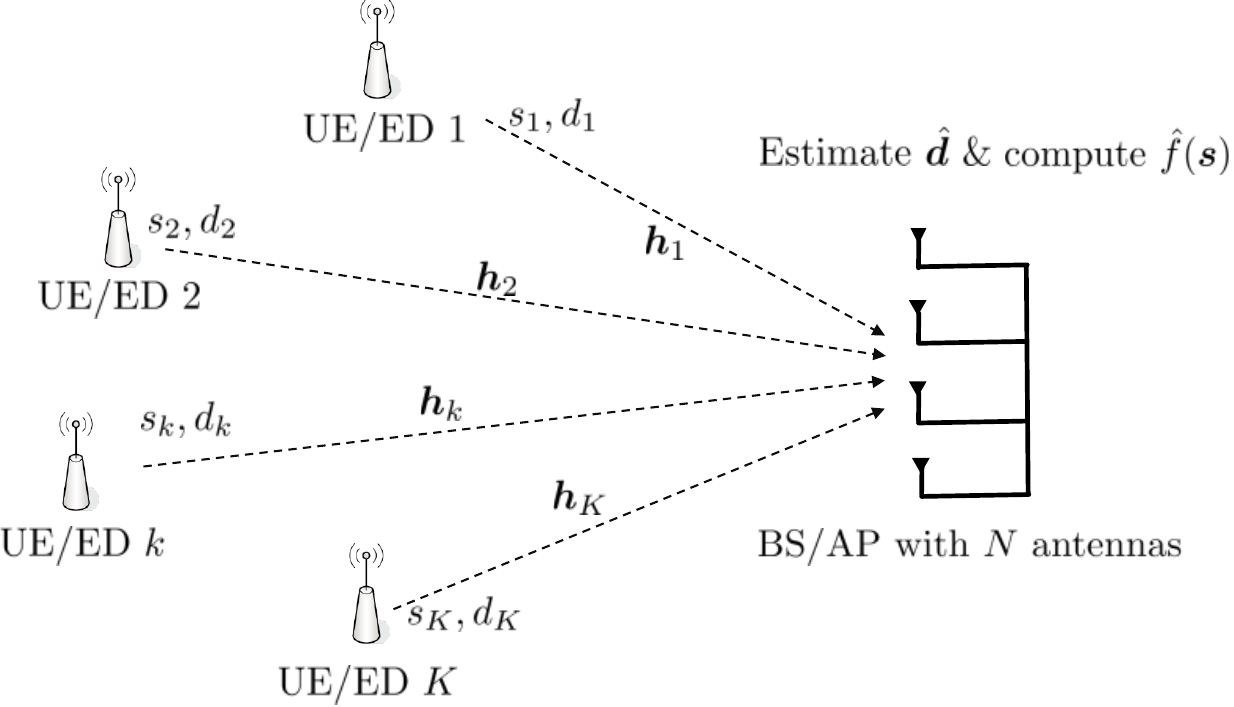}
\vspace{-4ex}
\caption{\ac{SIMO} \ac{ICC} system consisting of one \ac{BS}/\ac{AP} with $N$ antennas and $K$ single antenna \acp{UE}/\acp{ED}.}
\label{fig:system_model}
\vspace{-1ex}
\end{figure}

Consider a \ac{SIMO}\footnote{The extension to multi-user \ac{MIMO} systems, while straighforward, is also laborious as it requires the joint design of precoders, and therefore will be addressed in a follow-up article.} uplink setup composed of $K$ single-antenna \acp{UE}/\acp{ED} and one \ac{BS}/\ac{AP} equipped with $N$ antennas, as illustrated in Fig. \ref{fig:system_model}.

\vspace{-2ex}
\subsection{Uplink ICC Signal Model}

Under the assumption of perfect symbol synchronization amongst users, the received signal $\bm{y} \in \mathbb{C}^{N\times 1}$ at the \ac{BS}/\ac{AP} subjected to fading and noise is given by
\vspace{-0.5ex}
\begin{equation}
\label{eq:received_signal}
\bm{y} = \sum_{k=1}^K \bm{h}_k  {x}_k + \bm{w},
\vspace{-0.5ex}
\end{equation}
where ${x}_k\in \mathbb{C}$ is a multifunctional transmit signal from the $k$-th user, $\bm{w} \in \mathbb{C}^{N\times1}\sim \mathcal{CN}(0,\sigma^2_w\mathbf{I}_N)$ is the \ac{AWGN} vector, and $\bm{h}_k \in \mathbb{C}^{N \times 1}$ is the \ac{SIMO} channel vector of the $k$-th user to the \ac{BS}/\ac{AP} following the uncorrelated block Rayleigh fading model typically assumed in the \ac{OTAC} literature \cite{LiuTWC20,AndoCAMSAP2023}, such that the $(n,k)$-th elements $h_{n,k}\sim \mathcal{CN}(0,1)$ of the channel matrix $\bm{H}$ are assumed to be \ac{iid} circularly symmetric complex Gaussian random variable with zero mean and unit variance, and sufficient coherence time.

Under the \ac{ICC} paradigm, the transmit signal is decomposed as a sum of communication and computing components, $i.e.$
\begin{equation}
\label{eq:transmit_sig_decomposition}
x_k \triangleq d_k +  \psi_k(s_k),
\end{equation}
where $d_k \in \mathcal{D}$ and $s_k \in \mathbb{C}$ denote $k$-th user's modulated symbol for communication and computing, respectively, with $\mathcal{D}$ representing an arbitrary discrete constellation of cardinality $D$, $e.g.$ \ac{QAM}; while $\psi_k(\cdot)$ denotes the pre-processing function for \ac{OTAC} to be elaborated in the following section.

For future convenience, the received signal can now be reformulated in terms of matrices as
\begin{equation}
\bm{y} =  \bm{H}  \bm{x} + \bm{w} =\bm{H}  (\bm{d} + \bm{s}) + \bm{w},
\label{eq:received_signal_matrix}
\end{equation}
where the complex channel matrix $\bm{H} \triangleq [\bm{h}_1,\dots, \bm{h}_K] \in \mathbb{C}^{N\times K}$, the concatenated transmit signal $\bm{x} \triangleq [x_1,\dots, x_K]\trans \in \mathbb{C}^{K\times1}$, the data signal vector $\bm{d} \triangleq [d_1,\dots, d_K]\trans \in \mathcal{D}^{K \times 1} \subset \mathbb{C}^{K\times1}$ and the computing signal vector $\bm{s} \triangleq [\psi_1(s_1),\dots, \psi_K(s_K)]\trans \in \mathbb{C}^{K\times1}$ are explicitly defined.
\vspace{-2ex}

\subsection{Description of the AirComp Operation}

The \ac{OTAC} operation consists of the evaluation of a target function $f(\bm{s})$ at the \ac{BS}/\ac{AP}, which can be described as \cite{LiuTWC20}
\vspace{-1ex}
\begin{equation}
\label{eq:target_function_def}
f(\bm{s}) = \phi\bigg(\sum_{k=1}^K \psi_k(s_k)\bigg),
\vspace{-1ex}
\end{equation}
where $\phi$ represents the \ac{OTAC} post-processing function for a general nomographic expression.

Two classic examples of nomographic functions often considered in the \ac{OTAC} literature \cite{NazerTIT07,Wang_arxiv_2024} are as follows.

\vspace{0.5ex}
\subsubsection{The Arithmetic Sum Operation}

One of the simplest nomographic functions is the arithmetic sum operation given by
\vspace{-1ex}
\begin{equation}
\label{eq:target_function_def_SUM}
f^{\mathrm{SUM}}(\bm{s}) = \phi^{\mathrm{SUM}}\bigg(\sum_{k=1}^K \psi^{\mathrm{SUM}}_k(s_k)\bigg) = \sum_{k=1}^K s_k,
\vspace{-1ex}
\end{equation}
where the corresponding pre- and post-processing functions are defined as $\psi^{\mathrm{SUM}}_k(s_k) \triangleq s_k$ and $\phi^{\mathrm{SUM}}\big(\sum_{k=1}^K \psi_k(s_k)\big) \triangleq \sum_{k=1}^K \psi_k(s_k)$.

\vspace{0.5ex}
\subsubsection{The Arithmetic Product Operation}

Similarly to the above, we may define the pre- and post-processing functions as $\psi^{\mathrm{PROD}}_k(s_k) \triangleq \log_2(s_k)$ and $\phi^{\mathrm{PROD}}\big(\sum_{k=1}^K \psi_k(s_k)\big) \triangleq 2^{\left(\sum_{k=1}^K \psi_k(s_k)\right)}$ and obtain the product of the signals as
\vspace{-1ex}
\begin{equation}
\label{eq:target_function_def_PROD}
f^{\mathrm{PROD}}(\bm{s}) = \phi^{\mathrm{PROD}}\bigg(\sum_{k=1}^K \psi^{\mathrm{PROD}}_k(s_k)\bigg) = \prod_{k=1}^K s_k,
\vspace{-1ex}
\end{equation}
where the base 2 is chosen \ac{wlg}\footnote{It is trivial to see that the operation persists regardless of the chosen logarithmic base.} for ease of implementation in digital systems.

In the remainder of this article, unless otherwise specified, the arithmetic sum operation is chosen for the target function, $i.e., f(\bm{s}) = f^{\mathrm{SUM}}(\bm{s})$.
This choice is \ac{wlg}, but convenient and in favor of clarity of exposition, since a cumbersome power scaling has to be taken into account for the incorporation of the product operation \cite{LiuTWC20}.

\vspace{-3ex}
\subsection{A Note on Transmit Power Allocation}
\label{subsec:power_allocation}

Taking into account that data symbols are to be detected individually, while the computing stream is an aggregate function, we shall for the sake of fairness assume that each transmit data signal, and each aggregate computing stream is allocated the same power.
As a consequence, denoting the average powers allocated to a data symbol and a computing signal respectively by $E_\mathrm{D}$ and $E_\mathrm{S}$, we have
\vspace{-1ex}
\begin{equation}
\label{eq:power_allocation}
E_\mathrm{S} = \frac{E_\mathrm{D}}{K}.
\vspace{-1ex}
\end{equation}

As an example, normalizing the data symbols to a unit average power, such that $E_\mathrm{D} = 1$, in a system with $K = 100$ \acp{ED} engaged in the \ac{OTAC} operation, the average power of the computing signal transmitted by each \ac{ED} is $E_\mathrm{S} = 0.01$.
%

\vspace{-2ex}
\section{Benchmarking Scheme for ICC Systems}
\label{sec:benchmarking_scheme}

In this section, we describe a \ac{GaBP}-based joint data and computing signal detection scheme which will serve as a benchmark for the \ac{ICC} method proposed subsequently.
It will be assumed that communication symbols are modulated via \ac{QPSK}, while the computing signal of each $k$-th user follows $s_k\sim\mathcal{CN}(\mu_s,\sigma_s^2)$, with the variance $\sigma_s^2$ known, but the mean $\mu_s$ unknown 
to the receiver.

In the sequel, we will derive the message-passing rules for the joint estimation of data symbols and computing signals, yielding the corresponding estimated data vector $\hat{\bm{d}} \in \mathbb{C}^{K \times 1}$ and target function $\hat{f}(\bm{s}) \in \mathbb{C}$.
We point out that in this first scheme the general element-wise structure of the \ac{GaBP} will be leveraged to enable the estimation of \ul{the individual elements} of both the data symbol vector $\bm{d}$ and computing signal vector $\bm{s}$, although the estimation of $\bm{s}$ is not necessary to execute the \ac{OTAC} operation; and highlight that the latter is done only \ul{for the sake of benchmarking}, since in this case the Bayes-optimal denoisers for the best decoding/computing performance of corresponding estimate vector can be designed.

\vspace{-2ex}
\subsection{Joint Detection and Computing}

In order to derive the scalar \ac{GaBP} rules, let us first express equation \eqref{eq:received_signal_matrix} in an element-wise manner, $i.e.$
\begin{equation}
y_n = \sum_{k=1}^K h_{n,k} \cdot d_k + \sum_{k=1}^K h_{n,k} \cdot s_k + w_n.
\label{eq:received_signal_elementwise}
\end{equation}

Next, consider the $i$-th iteration of the message passing algorithm, and denote the soft replicas of the $k$-th communication and computing symbol with the $n$-th receive symbol $y_n$ at the previous iteration respectively by $\hat{d}_{n,k}^{(i-1)}$ and $\hat{s}_{n,k}^{(i-1)}$. 
Then, the \acp{MSE} of the soft-replicas computed for the $i$-th iteration are respectively given by
\begin{subequations}
\begin{eqnarray}
&\hat{\sigma}^{2(i)}_{d:{n,k}}\!  \triangleq\! \mathbb{E}_{d} \big[ | d\! -\! \hat{d}_{n,k}^{(i-1)} |^2 \big]\!=\! E_\mathrm{D}\! -\! |\hat{d}_{n,k}^{(i-1)}|^2,&
\label{eq:MSE_d_k}\\[1ex]
&\hat{\sigma}^{2(i)}_{s:{n,k}}  \triangleq \mathbb{E}_{s} \big[ | s - \hat{s}_{n,k}^{(i-1)} |^2 \big]\!=\! E_\mathrm{S}\! -\! |\hat{s}_{n,k}^{(i-1)}|^2,&
\label{eq:MSE_s_k}
\end{eqnarray}
\end{subequations}
$\forall (n,k)$, where $\mathbb{E}_{d}$ refers to expectation over all the possible symbols in the constellation $\mathcal{D}$, with average power $E_\mathrm{D}$, while $\mathbb{E}_{s}$ refers to expectation over all the possible outcomes of $s \sim \mathcal{CN}(\mu_s,\sigma^2_s)$, respectively.

\vspace{1ex}
\subsubsection{\Ac{sIC}} The objective of the \ac{sIC} stage is to compute, at a given $i$-th iteration of the algorithm, the data- and computing-centric \ac{sIC} symbols $\tilde{y}_{d:n,k}^{(i)}$ and $\tilde{y}_{s:n,k}^{(i)}$, as well as their corresponding variances $\tilde{\sigma}_{d:n,k}^{2(i)}$ and $\tilde{\sigma}_{s:n,k}^{2(i)}$, utilizing the soft replicas $\hat{d}_{n,k}^{(i-1)}$ and $\hat{s}_{n,k}^{(i-1)}$ from a previous iteration.
Exploiting equation \eqref{eq:received_signal_elementwise}, the \ac{sIC} data symbols and computing signals are given by
\begin{subequations}
\begin{equation}
\label{eq:d_soft_IC}
\resizebox{0.48 \textwidth}{!}{$
\begin{aligned}
\tilde{y}_{d:n,k}^{(i)} &= y_{n} - \sum_{q \neq k} h_{n,q}\hat{d}_{n,q}^{(i-1)} - \sum_{k=1}^K h_{n,k}\hat{s}_{n,k}^{(i-1)}, \\[-2ex]
&= h_{n,k} d_k + \underbrace{\sum_{q \neq k} h_{n,q}(d_q - \hat{d}_{n,q}^{(i-1)}) + \sum_{k = 1}^{K} h_{n,k}(s_k - \hat{s}_{n,k}^{(i-1)})+ w_n}_\text{interference + noise term}, \\[-12ex]
\end{aligned}
$}
\end{equation}
\vspace{6ex}
\begin{equation}
\label{eq:s_soft_IC}
\resizebox{0.48 \textwidth}{!}{$
\begin{aligned}
\tilde{y}_{s:n,k}^{(i)} &= y_{n} - \sum_{q \neq k} h_{n,q}\hat{s}_{n,q}^{(i-1)} - \sum_{k=1}^K h_{n,k}\hat{d}_{n,k}^{(i-1)}, \\[-2ex]
&= h_{n,k} s_k + \underbrace{\sum_{q \neq k} h_{n,q}(s_q - \hat{s}_{n,q}^{(i-1)}) + \sum_{k = 1}^{K} h_{n,k}(d_k - \hat{d}_{n,k}^{(i-1)})+ w_n}_\text{interference + noise term}.
\end{aligned}
$}
\end{equation}
\end{subequations}

\begin{subequations}
In turn, leveraging \ac{SGA} to approximate the interference and noise terms as Gaussian noise, the conditional \acp{PDF} of the \ac{sIC} symbols are given by
\vspace{-1ex}
\begin{eqnarray}
\label{eq:cond_PDF_d}
&\mathbb{P}_{\tilde{\mathrm{y}}_{\mathrm{d}:n,k}^{(i)} \mid \mathrm{d}_{k}}(\tilde{y}_{d:n,k}^{(i)}|d_{k}) \propto \mathrm{exp}\bigg[\! -\tfrac{|\tilde{y}_{d:n,k}^{(i)}\! -\! h_{n,k} d_{k}|^2}{\tilde{\sigma}_{d:n,k}^{2(i)}} \bigg],&\\
%
\label{eq:cond_PDF_s}
&\mathbb{P}_{\tilde{\mathrm{y}}_{\mathrm{s}:n,k}^{(i)} \mid \mathrm{s}_{k}}(\tilde{y}_{s:n,k}^{(i)}|s_{k}) \propto \mathrm{exp}\bigg[\! -\tfrac{|\tilde{y}_{s:n,k}^{(i)}\! -\! h_{n,k} s_{k}|^2}{\tilde{\sigma}_{s:n,k}^{2(i)}} \bigg],&
\end{eqnarray}
\end{subequations}
with corresponding conditional \ac{sIC} variances given by
\begin{subequations}
\begin{eqnarray}
\label{eq:soft_IC_var_d}
&\tilde{\sigma}_{d:n,k}^{2(i)}\! =\! \sum\limits_{q \neq k} \left|h_{n,q}\right|^2 \hat{\sigma}^{2(i)}_{d:{n,q}}\! +\! \sum\limits_{k = 1}^{K} \left|h_{n,k}\right|^2 \hat{\sigma}^{2(i)}_{s:{n,k}} + \sigma^2_w,&\\
\label{eq:soft_IC_var_s}
&\tilde{\sigma}_{s:n,k}^{2(i)}\! =\! \sum\limits_{q \neq k} \left|h_{n,q}\right|^2 \hat{\sigma}^{2(i)}_{s:{n,q}}\! +\! \sum\limits_{k = 1}^{K} \left|h_{n,k}\right|^2 \hat{\sigma}^{2(i)}_{d:{n,k}}\! +\! \sigma^2_w.&
\end{eqnarray}
\end{subequations}

\subsubsection{Belief Generation} With the goal of generating the beliefs for all the data and computing symbols, we first exploit \ac{SGA} under the assumption that $N$ is a sufficiently large number and that the individual estimation errors in $\hat{d}_{n,k}^{(i-1)}$ and $\hat{s}_{n,k}^{(i-1)}$ are independent.
Then, in hand of the conditional \acp{PDF}, the extrinsic \acp{PDF} of the $k$-th estimated data symbol and $k$-th estimated computing signal, respectively, are obtained as
\begin{subequations}
\begin{eqnarray}
\label{eq:extrinsic_PDF_d}
&\prod\limits_{q \neq n} \mathbb{P}_{\tilde{\mathrm{y}}_{\mathrm{d}:q,k}^{(i)} \mid \mathrm{d}_{k}}(\tilde{y}_{d:q,k}^{(i)}|d_{k}) \propto \mathrm{exp}\bigg[ - \tfrac{(d_k - \bar{d}_{n,k}^{(i)})^2}{\bar{\sigma}_{d:n,k}^{2(i)}} \bigg],&\\
\label{eq:extrinsic_PDF_s}
&\prod\limits_{q \neq n} \mathbb{P}_{\tilde{\mathrm{y}}_{\mathrm{s}:q,k}^{(i)} \mid \mathrm{s}_{k}}(\tilde{y}_{s:q,k}^{(i)}|s_{k}) \propto \mathrm{exp}\bigg[ - \tfrac{(s_k - \bar{s}_{n,k}^{(i)})^2}{\bar{\sigma}_{s:n,k}^{2(i)}} \bigg],&
\end{eqnarray}
\end{subequations}
where the corresponding extrinsic means are defined as
\begin{subequations}
\begin{eqnarray}
\label{eq:extrinsic_mean_d}
&\bar{d}_{n,k}^{(i)} \triangleq \bar{\sigma}_{d:n,k}^{(i)} \sum_{q \neq n} \frac{h^*_{q,k} \cdot \tilde{y}_{d:q,k}^{(i)}}{ \tilde{\sigma}_{d:q,k}^{2(i)}},&\\
\label{eq:extrinsic_mean_s}
&\bar{s}_{n,k}^{(i)} \triangleq \bar{\sigma}_{s:n,k}^{(i)}  \sum_{q \neq n} \frac{h^*_{q,k} \cdot \tilde{y}_{s:q,k}^{(i)}}{ \tilde{\sigma}_{s:q,k}^{2(i)}},&
\end{eqnarray}
\end{subequations}
with the extrinsic variances given by
\begin{subequations}
\begin{eqnarray}
\label{eq:extrinsic_var_d}
&\bar{\sigma}_{d:n,k}^{2(i)} = \bigg( \sum_{q \neq n} \frac{|h_{q,k}|^2}{\tilde{\sigma}_{d:q,k}^{2(i)}} \bigg)^{\!\!\!-1},&\\
\label{eq:extrinsic_var_s}
&\bar{\sigma}_{s:n,k}^{2(i)} = \bigg( \sum_{q \neq n} \frac{|h_{q,k}|^2}{\tilde{\sigma}_{s:q,k}^{2(i)}} \bigg)^{\!\!\!-1}.&
\end{eqnarray}
\end{subequations}

\subsubsection{Soft Replica Generation} This stage involves the exploitation of the previously computed beliefs and their denoising via a Bayes-optimal denoiser which yield the final estimates for the intended variables.
A damping procedure can also be incorporated here to prevent convergence to local minima due to incorrect hard-decision replicas.

Under the assumption that the communication data symbols are taken from a \ac{QPSK} constellation, the Bayes-optimal denoiser is given\footnote{For other types of modulation, the corresponding alternative denoisers are required, but can be easily derived \cite{takahashi2018design}.} by
\begin{equation}
\vspace{-1ex}
\hat{d}_{n,k}^{(i)}\! =\! c_d\! \cdot\! \bigg(\! \text{tanh}\!\bigg[ 2c_d \frac{\Real{\bar{d}_{n,k}^{(i)}}}{\bar{\sigma}_{d:{n,k}}^{2(i)}} \bigg]\!\! +\! j\text{tanh}\!\bigg[ 2c_d \frac{\Imag{\bar{d}_{n,k}^{(i)}}}{\bar{\sigma}_{{d}:{n,k}}^{2(i)}} \bigg]\!\bigg),\!\!
\label{eq:QPSK_denoiser}
\end{equation}
where $c_d \triangleq \sqrt{E_\mathrm{D}/2}$ denotes the magnitude of the real and imaginary parts of the explicitly chosen \ac{QPSK} symbols, with its corresponding variance updated as in equation \eqref{eq:MSE_d_k}.

Similarly, since the computing signal follows a Gaussian distribution, the denoiser with a Gaussian prior and its corresponding variance are given by~\cite{ItoTCOM_2024}
\begin{subequations}
\begin{equation}
\label{eq:s_denoiser_mean}
\hat{s}_{n,k}^{(i)} = \frac{\sigma^2_s \cdot \bar{s}_{n,k}^{(i)} + \bar{\sigma}_{s:n,k}^{2(i)} \cdot \hat{\mu}_s^{(i)}}{\bar{\sigma}_{s:n,k}^{2(i)} + \sigma^2_s},
\end{equation}
\begin{equation}
\label{eq:s_denoiser_var}
\hat{\sigma}_{s:n,k}^{2(i)} = \frac{\sigma^2_s \cdot \bar{\sigma}_{s:n,k}^{2(i)}}{\bar{\sigma}_{s:n,k}^{2(i)} + \sigma^2_s}.
\end{equation}
\end{subequations}
where $\hat{\mu}_s^{(i)}$ represents an estimate of the true mean of $s_k$, and in the first iteration, it should be properly initialized, \textit{i.e.}, $\hat{\mu}_s^{(0)} = 0$ when no prior information is available.

From the second iteration onward, $\hat{\mu}_s^{(i)}$ is updated according to the \ac{EM} algorithm, as detailed in Section \ref{subsubsec:EM_alg}.
After obtaining $\hat{d}_{n,k}^{(i)}$ and $\hat{s}_{n,k}^{(i)}$ as per equations \eqref{eq:QPSK_denoiser} and \eqref{eq:s_denoiser_mean}, the final outputs are computed by damping the results with damping factors $0 < \beta_d$ and $\beta_s < 1$ in order to improve convergence \cite{Su_TSP_2015}, yielding
\begin{subequations}
\begin{equation}
\label{eq:d_damped}
\hat{d}_{n,k}^{(i)} = \beta_d \hat{d}_{n,k}^{(i)} + (1 - \beta_d) \hat{d}_{n,k}^{(i-1)},
\end{equation}
\begin{equation}
\label{eq:s_damped}
\hat{s}_{n,k}^{(i)} = \beta_s \hat{s}_{n,k}^{(i)} + (1 - \beta_s) \hat{s}_{n,k}^{(i-1)}.
\end{equation}
\end{subequations}

In turn, the corresponding variances $\hat{\sigma}^{2(i)}_{d:{n,k}}$ and $\hat{\sigma}^{2(i)}_{s:{n,k}}$ are first updated via equations \eqref{eq:MSE_d_k} and \eqref{eq:s_denoiser_var}, respectively, and then damped via
\begin{subequations}
\begin{equation}
\label{eq:MSE_d_m_damped}
\hat{\sigma}^{2(i)}_{d:{n,k}} = \beta_d \hat{\sigma}_{d:{n,k}}^{2(i)} + (1-\beta_d) \hat{\sigma}_{d:{n,k}}^{2(i-1)},
\end{equation}
\begin{equation}
\label{eq:MSE_s_m_damped}
\hat{\sigma}^{2(i)}_{s:{n,k}} = \beta_s \hat{\sigma}_{s:{n,k}}^{2(i)} + (1-\beta_s) \hat{\sigma}_{s:{n,k}}^{2(i-1)}.
\end{equation}
\end{subequations}

Finally, the consensus update can be obtained as
\begin{subequations}
\begin{equation}
\label{eq:d_hat_final_est}
\hat{d}_{k} = \bigg( \sum_{n=1}^N \frac{|h_{n,k}|^2}{\tilde{\sigma}_{d:n,k}^{2(i)}} \bigg)^{\!\!\!-1} \! \! \bigg( \sum_{n=1}^N \frac{h^*_{n,k} \cdot \tilde{y}_{d:n,k}^{(i)}}{ \tilde{\sigma}_{d:n,k}^{2(i)}} \bigg),
\end{equation}
\begin{equation}
\label{eq:s_hat_final_est}
\hat{s}_{k} =  \bigg( \sum_{n=1}^N \frac{|h_{n,k}|^2}{\tilde{\sigma}_{s:n,k}^{2(i)}} \bigg)^{\!\!\!-1} \! \! \bigg( \sum_{n=1}^N \frac{h^*_{n,k} \cdot \tilde{y}_{s:n,k}^{(i)}}{ \tilde{\sigma}_{s:n,k}^{2(i)}} \bigg).
\end{equation}
\end{subequations}

\subsubsection{Expectation Maximization Update}
\label{subsubsec:EM_alg}

The estimate of the true mean $\hat{\mu}_s$ can be updated iteratively via the \ac{EM} algorithm.
Since a detailed derivation of the \ac{EM} procedure exploiting the Kullback-Leibler divergence and log likelihood can be found in \cite{ItoICC2023,RanasingheARXIV2024}, it suffices to state that the update rule for Gaussian-distributed variables can be expressed as
\begin{equation}
\label{eq:EM_update}
\hat{\mu}_s^{(i)} = \frac{1}{K} \sum_{k=1}^K \hat{s}_k^{(i)}.
\end{equation}

\subsection{Closed-form AirComp Combiner Design}

In order to faciliate the actual computation of the target function at the \ac{BS}/\ac{AP}, let us first write the combining of the residual signal leveraging equation \eqref{eq:received_signal_matrix} after \ac{SIC} of the estimated communication signal $\hat{\bm{d}}$ as\footnote{If $s_k \in \mathbb{R}$, the real part of $\hat{f}(\bm{s})$ can be extracted from equation \eqref{eq:Aircomp_SIC}.}
\vspace{-1ex}
\begin{equation}
\vspace{-1ex}
\label{eq:Aircomp_SIC}
\hat{f}(\bm{s}) = \bm{u}\herm(\bm{y} - \bm{H}\hat{\bm{d}}) = \bm{u}\herm(\bm{H} (\bm{s} - \check{\bm{d}}) +\bm{w}),
\end{equation}
where $\bm{u} \in \mathbb{C}^{N\times1}$ denotes the combining vector, and we intrinsically define a data signal error vector $\check{\bm{d}} \triangleq \hat{\bm{d}} - \bm{d} \in \mathbb{C}^{K \times 1}$.

Leveraging the above formulation, let us consider the optimization problem given by \vspace{-0.5ex}
\begin{equation}
\label{eq:min_problem}
\underset{\bm{u}\in\mathbb{C}^{N\times1}}{\mathrm{minimize}} \hspace{3ex} \mathbb{E} \big[ \| f(\bm{s}) - \hat{f}(\bm{s})  \|_2^2 \big], \vspace{-0.5ex}
\end{equation}
where the objective function is defined as
\begin{align}
\label{eq:obj_func_def}
\| f(\bm{s}) \!-\! \hat{f}(\bm{s})  \|_2^2 \!\triangleq\! \| \mathbf{1}_{K}\trans \bm{s} \!-\! \bm{u}\herm(\bm{H} (\bm{s} \!-\! \check{\bm{d}}) +\bm{w})  \|_2^2.
\end{align}

Then, the closed-form solution for the combining vector can be derived as
\begin{equation}
\label{eq:u_def_precoder}
\bm{u} = (\bm{H}(\sigma_s^2\mathbf{I}_{K} + \boldsymbol{\Omega} )\bm{H}\herm+\sigma^2_w \mathbf{I}_{N})^{-1} \bm{H} \sigma_s^2 \mathbf{1}_{K},
\end{equation}
where $\boldsymbol{\Omega} \triangleq  \mathbb{E} [\check{\bm{d}} \check{\bm{d}}\herm] = (\hat{\bm{\Sigma}}^{(i_\text{max})}_{d})\herm (\hat{\bm{\Sigma}}^{(i_\text{max})}_{d})$, with $\hat{\bm{\Sigma}}^{(i_\text{max})}_{d}$ denoting the matrix collecting the standard deviations $\hat{\sigma}^{(i_\text{max})}_{d:{n,k}}$ for all symbol estimation errors, computed from equation \eqref{eq:MSE_d_k} at the final iteration of the \ac{GaBP} algorithm.

\vspace{-2ex}
\subsection{Joint Integrated Communication and Computing Design}

We now combine the low complexity \ac{GaBP} with the closed-form \ac{OTAC} combiner to estimate both the data signal $\hat{\bm{d}}$ and obtain the computing function $\hat{f}(\bm{s})$.
The complete pseudocode for the procedure is summarized in Algorithm \ref{alg:proposed_ICC}.
\vspace{-1ex}
\begin{algorithm}[H]
\caption{Benchmarking Joint Data Detection \& \ac{OTAC} for Integrated  Communication and Computing Systems}
\label{alg:proposed_ICC}
\setlength{\baselineskip}{11pt}
\textbf{Input:} receive signal vector $\bm{y}\in\mathbb{C}^{N\times 1}$, complex channel matrix $\bm{H}\in\mathbb{C}^{N\times K}$, maximum number of iterations $i_{\max}$, data constellation power $E_\mathrm{D}$, computing signal variance $\sigma^2_s$, noise variance $\sigma^2_w$ and damping factors $\beta_d,\beta_s$. \\
\textbf{Output:} $\hat{\bm{d}}$, $\hat{\bm{s}}$ and $\hat{f}(\bm{s})$ 
\vspace{-2ex} 
\begin{algorithmic}[1]  
\STATEx \hspace{-3.5ex}\hrulefill
\STATEx \hspace{-3.5ex}\textbf{Initialization}
\STATEx \hspace{-3.5ex} - Set iteration counter to $i=0$ and amplitudes $c_d = \sqrt{E_\mathrm{D}/2}$.
\STATEx \hspace{-3.5ex} - Set initial data estimates to $\hat{d}_{n,k}^{(0)} = 0$
%
%
and corresponding 
\STATEx \hspace{-2ex} variances to $\hat{\sigma}^{2(0)}_{d:{n,k}} = E_\mathrm{D}, \forall n,k$.
\STATEx \hspace{-3.5ex} - Set initial computing signal estimates to $\hat{s}_{n,k}^{(0)} = 0$ and 
\STATEx \hspace{-2ex} corresponding variances to $\hat{\sigma}^{2(0)}_{s:{n,k}} = \sigma_s^2, \forall n, k$.
\STATEx \hspace{-3.5ex} - Set $\hat{\mu}_s^{(0)} = 0$.
\STATEx \hspace{-3.5ex}\hrulefill
\STATEx \hspace{-3.5ex}\textbf{for} $i=1$ to $i_\text{max}$ \textbf{do}
\STATEx \textbf{Communication and Computing Update}: $\forall n, k$
\STATE Compute \ac{sIC} data signal $\tilde{y}_{d:{n,k}}^{(i)}$ and its corresponding variance $\tilde{\sigma}^{2(i)}_{d:{n,k}}$ from equations \eqref{eq:d_soft_IC} and \eqref{eq:soft_IC_var_d}.
\STATE Compute \ac{sIC} computing signal $\tilde{y}_{s:{n,k}}^{(i)}$ and its corresponding variance $\tilde{\sigma}^{2(i)}_{s:{n,k}}$ from equations \eqref{eq:s_soft_IC} and \eqref{eq:soft_IC_var_s}.
\STATE Compute extrinsic data signal belief $\bar{d}_{n,k}^{(i)}$ and its corresponding variance $\bar{\sigma}_{d:{n,k}}^{2(i)}$ from equations \eqref{eq:extrinsic_mean_d} and \eqref{eq:extrinsic_var_d}.
\STATE Compute extrinsic computing signal belief $\bar{s}_{n,k}^{(i)}$ and its corresponding variance $\bar{\sigma}_{s:{n,k}}^{2(i)}$ from eqs. \eqref{eq:extrinsic_mean_s} and \eqref{eq:extrinsic_var_s}.
\STATE Compute denoised and damped data signal estimate $\hat{d}_{n,k}^{(i)}$ from equations \eqref{eq:QPSK_denoiser} and \eqref{eq:d_damped}.
\STATE Compute denoised and damped data signal variance $\hat{\sigma}_{d:{n,k}}^{2(i)}$ from equations \eqref{eq:MSE_d_k} and \eqref{eq:MSE_d_m_damped}.
\STATE Compute denoised and damped computing signal estimate $\hat{s}_{n,k}^{(i)}$ from equations \eqref{eq:s_denoiser_mean} and \eqref{eq:s_damped}.
\STATE Compute denoised and damped computing signal variance $\hat{\sigma}_{s:{n,k}}^{2(i)}$ from equations \eqref{eq:s_denoiser_var} and \eqref{eq:MSE_s_m_damped}.
\STATE Compute $\hat{s}_k^{(i)}, \forall k$ using equation \eqref{eq:s_hat_final_est}.
\STATE Update $\hat{\mu}_s^{(i)}$ using equation \eqref{eq:EM_update}.

\STATEx \hspace{-3.5ex}\textbf{end for}
\STATEx \hspace{-3.5ex}\textbf{Communication and Computing Consensus}: 
\STATE Calculate $\hat{d}_k, \forall k$ (equivalently $\hat{\bm{d}}$) using equation \eqref{eq:d_hat_final_est}. 
\STATE Compute $\bm{u}$ from equation \eqref{eq:u_def_precoder}.
\STATE Compute $\hat{f}(\bm{s})$ from equation \eqref{eq:Aircomp_SIC}.
\end{algorithmic}
\end{algorithm}
\newpage

\vspace{-2ex}
\subsection{Performance and Complexity Analysis}
\label{subsec:performance_analysis_benchmarking}

In order to evaluate the performance of Algorithm \ref{alg:proposed_ICC}, we consider a typical uplink system composed of a \ac{BS}/\ac{AP} with $N = 100$ antennas servicing a varying number of single-antenna users, namely $K = 75$, $K = 100$ and $K = 125$, resulting in underloaded, fully loaded and overloaded scenarios, respectively\footnote{Notice that as a result of the bivariate estimation carried out, even the case when $N=100$ and $K=75$ is already overloaded from an estimation theoretical point-of-view since a total of $2K$ variables in the form of data and computing symbols need to be estimated from $N$ factor nodes.
A potential solution to mitigate this issue would be to take advantage of a linear inference scheme with a Gaussian mixture \cite{VilaTSP2013} model to directly estimate $x_k, \forall k$ in equation \eqref{eq:received_signal}, bringing back the default loading structure.}.   

As a consequence of the system model described by equation \eqref{eq:received_signal_matrix}, we must distinguish between the \ac{SINR} affecting data detection and the \ac{SNR} affecting the \ac{OTAC} operation, hereafter denoted as $\text{SINR}_D$ and $\text{SNR}_S$, respectively,
In particular, the \ac{SINR} for data symbols is defined as
%
\begin{equation}
\label{eq:SINR_def}
\text{SINR}_D \triangleq \frac{\mathbb{E}[|| \bm{H} \bm{d} ||^2]}{\alpha_S \mathbb{E}[|| \bm{H} \bm{s} ||^2] + \sigma^2_w},
\end{equation}
where $\alpha_S \in \{0,1\}$ is a parameter introduced to account for whether interference cancellation is performed or not, such that $\alpha_S = 0$ if $\bm{s}$ is explicitly estimated and cancelled, as is the case under Algorithm \ref{alg:proposed_ICC}; while $\alpha_S = 1$ when only $f(\bm{s})$ is estimated, such that the computing symbols $\bm{s}$ remain unknown to the receiver and treated as interference in the estimation of the data vector $\bm{s}$, as is the case of the proposed methods to be introduced subsequently.

In turn, the \ac{SNR} for the computing signal is defined as
%
\begin{equation}
\label{eq:SINR_S_def}
\text{SNR}_S \triangleq \frac{\mathbb{E}[|| \bm{H} \bm{s} ||^2]}{\sigma^2_w}.
\end{equation}

In all simulations carried out, the total transmit power is distributed as described in subsection \ref{subsec:power_allocation}, with the computing signal  assumed to follow $s_k\sim\mathcal{CN}(0,E_\mathrm{S})$ and the channel coefficients following $h_{n,k} \sim \mathcal{CN}(0,1)$.
Since this scheme is to serve as a benchmark for the methods contributed hereafter, the factors $\alpha_D$ and $\alpha_S$ in equations \eqref{eq:SINR_def} and \eqref{eq:SINR_S_def} are both set to $0$, implicating that the estimation procedure iteratively cancels out the interference from the data and computing signals.
Finally, the algorithmic damping parameters and number of iterations are set as $\beta_d = 0.5$, $\beta_s = 0.8$ and $i_\text{max} = 30$.

In addition to the typical \ac{BER} used to assess the performance of communication systems, we also consider the \ac{NMSE} as the metric to evaluate the performance of the computing function estimation in order to draw fair comparisons between all the loading scenarios, defined as
\begin{equation}
\label{eq:MSE_def}
\text{NMSE} \triangleq \frac{|| f(\bm{s}) - \hat{f}(\bm{s}) ||^2_2}{K}.
\end{equation}

\begin{figure}[H]
\subfigure[{\footnotesize $N=100$}]%
{\includegraphics[width=\columnwidth]{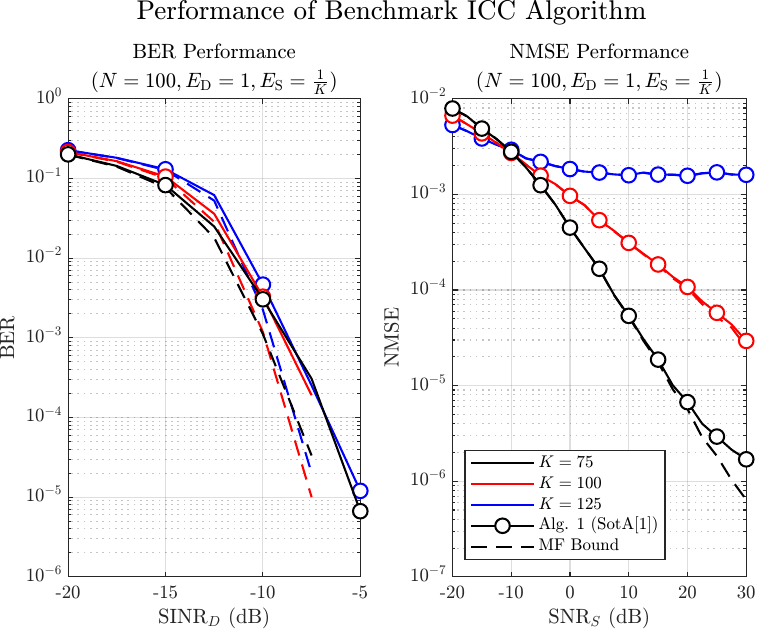}
\label{fig:BER_MSE_plot_1}}\\
\subfigure[{\footnotesize $N=200$}]%
{\includegraphics[width=\columnwidth]{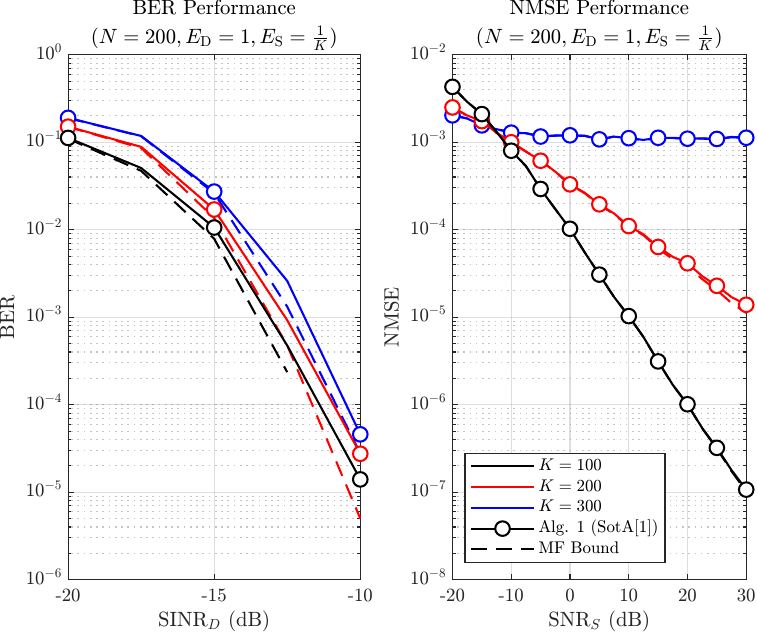}
\label{fig:BER_MSE_plot_2}}
\vspace{-2ex}
\caption{\ac{BER} and \ac{NMSE} achieved by the benchmark scheme  (Algorithm \ref{alg:proposed_ICC}) in overloaded, underloaded and fully-loaded scenarios, under two different system sizes, and with a single computing stream.}
\label{fig:BER_MSE_Alg1}
\vspace{-1ex}
\end{figure}

Our first results, shown in Fig. \ref{fig:BER_MSE_Alg1}, showcases the \ac{BER} and \ac{NMSE} associated with the estimates $\hat{\bm{d}}$ and $\hat{f}(\bm{s})$, respectively, for all three aforementioned scenarios, and systems of two different scales, determined by $N=100$ and $N=200$, respectively.

In addition to the curves obtained from full simulations, curves for the \ac{MF} matched filter bound performance are also shown, which correspond to the case when the algorithm is executed initialized by the true solution.

It can be seen from the results that the integration of \ac{OTAC} has little impact onto the \ac{GaBP}-based detection of digitally-modulated symbols, regardless of loading condition.
The converse, however, is not quite the same, since the performance of the \ac{OTAC} operation clearly degrades significantly as the loading condition worsens, regardless of system size.

At first sight, this result may be counterintuitive, since the weak impact of loading onto \ac{BER} implies that communications symbols are accurately detected, such that the corresponding interference can be effectively removed, which in turn would suggest that the performance of estimating $f(\bm{s})$ should not depend on $K$.
In hindsight, however, this arises from the fundamental fact that the data symbols are subject to discrete constraints, whereas the computing stream is a continuous signal. 
Since discrete constraints translate to a form of sparsity, accurate estimation is still possible even in overloaded scenarios. 
In contrast, estimating continuous signals becomes a severely ill-conditioned problem, making it difficult to achieve good performance under overload.
Furthermore, another reason is a consequence of the power allocation among the data and computing stream, since significantly less power is allocated to computing as the number of communications users grows.

In any case, it is worth noting that the while the performance of the \ac{OTAC} operation in \ac{ICC} under overloaded conditions reaches the corresponding Matched-Filer bound, the overall performance is still poor, as a result of the combiner described by equation \eqref{eq:u_def_precoder}, which is not effective\footnote{Notice that for large $K$, $\sigma^2_s \to 0$, and for large \ac{SINR}, both $\sigma^2_w \to 0$ and $\bm{\Omega} \to \bm{0}$, such that the combining beamformer $\bm{u}$ becomes ineffective \cite{AndoCAMSAP2023}.} when $K>N$ at large \acp{SINR}.
We therefore will no longer consider the overloading scenario hereafter.

Finally, we draw attention to the fact that, thanks to the low complexity of the \ac{GaBP} framework, systems of quite large size can also be practically considered.
In particular, except for the cubic complexity order -- namely $\mathcal{O}(N^3)$ -- of the last \ac{OTAC} combining step (line 13) of Algorithm \eqref{alg:proposed_ICC}, the repetitive steps of the algorithm amount to a complexity order of $\mathcal{O}(NK)$.

%
%

\section{Proposed Single-Stream ICC Framework}
\label{sec:proposed_method}

While it was shown in the previous section that the \ac{GaBP} algorithm can be used to jointly detect the data and computing signals in a benchmarking fashion, two main issues persist.
First, since the scheme described in Algorithm \ref{alg:proposed_ICC} estimates all the computing signals $s_k, \forall k$ separately, the fundamental \ac{OTAC} operation -- namely the computation of $f(\bm{s})$ \ul{over-the-air} -- is not truly carried out.
Second, the computation of the combiner described in line 12 of Algorithm \ref{alg:proposed_ICC} has cubic complexity order, which becomes prohibitive in large systems.

Leveraging the fact that the fundamental \ac{OTAC} operation resumes to a \ul{combiner design} problem, as shown by related literature \cite{QinWCL21,AndoCAMSAP2023}, we present a novel single-stream \ac{ICC} framework based on the \ac{GaBP} algorithm in which the matrix inversion in equation \eqref{eq:u_def_precoder} is removed, reducing the complexity of the scheme to $\mathcal{O}(N^2)$, with an intermediate low-rank combiner design that has complexity of order $\mathcal{O}(N^2K)$.
In addition, in contrast to the previously described benchmarking scheme in Section \ref{sec:benchmarking_scheme}, in this proposed framework, the desired target function $\hat{f}(\bm{s})$ is computed directly, as opposed to the individual computing signals in the vector $\hat{\bm{s}}$, such that actual \ac{ICC} is carried out.

To that end, let us revisit the signal model in eq. \eqref{eq:received_signal_matrix}, which can be reformulated as
\begin{equation}
\bm{y} = \bm{H}  \bm{d} + \underbrace{\bm{H}  \bm{s} + \bm{w}}_{\tilde{\bm{w}}},
\label{eq:received_signal_matrix_ICC_cov_noise}
\end{equation}
where $\tilde{\bm{w}} \triangleq \bm{H}  \bm{s} + \bm{w}$ will be considered as an effective noise vector for the purpose of data detection.

The effective noise covariance matrix of $\tilde{\bm{w}}$ is given by 
\vspace{-1ex}
\begin{align}
\vspace{-1ex}
\label{eq:combined_cov_matrix}
\bm{R}_{\tilde{\bm{w}}} \triangleq \mathbb{E}[\tilde{\bm{w}} \tilde{\bm{w}}\herm] &= \mathbb{E}[(\bm{H}  \bm{s} + \bm{w})(\bm{H}  \bm{s} + \bm{w})\herm] \nonumber \\
&= \mathbb{E}[(\bm{H}  \bm{s} + \bm{w}) (\bm{s}\herm \bm{H}\herm + \bm{w}\herm)] \nonumber \\
&= \mathbb{E}[ \bm{H}\bm{s}\bm{s}\herm \bm{H}\herm \!+\! \bm{H}\bm{s}\bm{w}\herm \!+\! \bm{w}\bm{s}\herm \bm{H}\herm \!+\! \bm{w}\bm{w}\herm ] \nonumber \\
&= \sigma_s^2 \underbrace{\bm{H}\bm{H}\herm}_{\bm{\Xi} \;\in\; \mathbb{C}^{N \times N}} + \sigma_w^2\mathbf{I}_N\nonumber\\
&\approx \sigma_s^2 \; \text{diag}(\bm{\xi})  + \sigma_w^2\mathbf{I}_N,
\end{align} 
where 
\vspace{-1ex}
\begin{equation}
\label{eq:channel_cov_vect}
\vspace{-1ex}
\bm{\xi} \triangleq [\xi_{1},\cdots,\xi_{N}]\trans \in \mathbb{C}^{N \times 1}.
\end{equation}
and the cross-terms in the expectation approach zero due to the zero-mean assumption over $\bm{w}$, and we implicitly defined the channel covariance matrix, with the last approximate equality extracted due to $i$) the assumption of \textit{channel hardening}, which holds when the size of $\bm{H}$ is sufficiently large, and $ii$) the fact that $\sigma_s^2$ is relatively small.

\vspace{-2ex}
\subsection{Linear GaBP Derivation for Data Detection}

Next, we derive the message-passing rules for the proposed single-stream \ac{ICC} method.
To that end, it will prove convenient to focus on a given $i$-th iteration of the algorithm, and denote the soft replicas of the $k$-th communication symbol with the $n$-th receive signal $y_n$ at the previous iteration by $\hat{d}_{n,k}^{(i-1)}$.
Then, the \acp{MSE} of these estimates computed for the $i$-th iteration are given by
\begin{equation}
\hat{\sigma}^{2(i)}_{d:{n,k}} \triangleq \mathbb{E}_{\mathrm{d}_{n,k}} \big[ | d_{n,k} - \hat{d}_{n,k}^{(i-1)} |^2 \big]= E_\mathrm{D} - |\hat{d}_{n,k}^{(i-1)}|^2,
\label{eq:MSE_d_k_P1}
\end{equation}
$\forall (n,k)$, where $\mathbb{E}_{\mathrm{d}_{n,k}}$ refers to the expectation over all the possible symbols in the constellation $\mathcal{D}$ with $E_\mathrm{D}$ explicitly denoting the data constellation power.

Then, the estimation of all the data symbols to fully recover the estimated data vector $\hat{\bm{d}} \in \mathbb{C}^{K \times 1}$ via the \ac{GaBP} technique can be carried out as follows.

\subsubsection{Soft Interference Cancellation} The objective of the \ac{sIC} stage at a given $i$-th iteration of the algorithm is to utilize the soft replicas $\hat{d}_{n,k}^{(i-1)}$ from a previous iteration to calculate the data-centric \ac{sIC} symbols $\tilde{y}_{d:n,k}^{(i)}$ with its corresponding variance $\tilde{\sigma}_{d:n,k}^{2(i)}$.
Therefore, exploiting equation \eqref{eq:received_signal_matrix_ICC_cov_noise}, the \ac{sIC} symbols for the data signals are given by
\begin{equation}
\label{eq:d_soft_IC_P1}
\begin{aligned}
\tilde{y}_{d:n,k}^{(i)} &= y_{n} - \sum_{q \neq k} h_{n,q}\hat{d}_{n,q}^{(i-1)} , \\
&= h_{n,k} d_k + \underbrace{\sum_{q \neq k} h_{n,q}(d_q - \hat{d}_{n,q}^{(i-1)}) + \tilde{w}_n}_\text{interference + noise term}. 
\end{aligned}
\end{equation}

In turn, leveraging \ac{SGA} to approximate the interference and noise terms as Gaussian noise, the conditional \ac{PDF} of the \ac{sIC} symbols are given by
\vspace{-1ex}
\begin{equation}
\label{eq:cond_PDF_d_P1}
\mathbb{P}_{\tilde{\mathrm{y}}_{\mathrm{d}:n,k}^{(i)} \mid \mathrm{d}_{k}}(\tilde{y}_{d:n,k}^{(i)}|d_{k}) \propto \mathrm{exp}\bigg[ -\frac{|\tilde{y}_{d:n,k}^{(i)} - h_{n,k} d_{k}|^2}{\tilde{\sigma}_{d:n,k}^{2(i)}} \bigg],
\vspace{-1ex}
\end{equation}


\noindent with its conditional variances expressed as
\begin{equation}
\label{eq:soft_IC_var_d_P1}
\tilde{\sigma}_{d:n,k}^{2(i)} = \sum_{q \neq k} \left|h_{n,q}\right|^2 \hat{\sigma}^{2(i)}_{d:{n,q}} + \sigma^2_{\tilde{w}:n},
\end{equation}
where, from equation \eqref{eq:combined_cov_matrix}, we have
\begin{equation}
\label{eq:noise_approximation_P1}
\sigma_{\tilde{w}:n}^2 \approx \sigma_s^2 \xi_n + \sigma_w^2.
\end{equation}

\subsubsection{Belief Generation} With the goal of generating the beliefs for all the data symbols, we first exploit \ac{SGA} under the assumption that $N$ is a sufficiently large number and that the individual estimation errors in $\hat{d}_{n,k}^{(i-1)}$ are independent.

Therefore, as a consequence of \ac{SGA} and in hand of the conditional \ac{PDF}, the extrinsic \ac{PDF} is obtained as
\begin{equation}
\label{eq:extrinsic_PDF_d_P1}
\prod_{q \neq n} \mathbb{P}_{\tilde{\mathrm{y}}_{\mathrm{d}:q,k}^{(i)} \mid \mathrm{d}_{k}}(\tilde{y}_{d:q,k}^{(i)}|d_{k}) \propto \mathrm{exp}\bigg[ - \frac{(d_k - \bar{d}_{n,k}^{(i)})^2}{\bar{\sigma}_{d:n,k}^{2(i)}} \bigg],
\end{equation}
where the corresponding extrinsic means are defined as
\begin{equation}
\label{eq:extrinsic_mean_d_P1}
\bar{d}_{n,k}^{(i)} = \bar{\sigma}_{d:n,k}^{(i)} \sum_{q \neq n} \frac{h^*_{q,k} \cdot \tilde{y}_{d:q,k}^{(i)}}{ \tilde{\sigma}_{d:q,k}^{2(i)}},
\end{equation}
with the extrinsic variances given by
\begin{equation}
\label{eq:extrinsic_var_d_P1}
\bar{\sigma}_{d:n,k}^{2(i)} = \bigg( \sum_{q \neq n} \frac{|h_{q,k}|^2}{\tilde{\sigma}_{d:q,k}^{2(i)}} \bigg)^{\!\!\!-1}.
\end{equation}

\subsubsection{Soft Replica Generation} This stage involves the exploitation of the previously computed beliefs and denoising them via a Bayes-optimal denoiser to get the final estimates for the intended variables.
A damping procedure can also be incorporated here to prevent convergence to local minima due to incorrect hard-decision replicas.

Since the data symbols originate from a discrete \ac{QPSK} alphabet, \ac{wlg}, the Bayes-optimal denoiser is given by
\begin{equation}
\hat{d}_{n,k}^{(i)}\! =\! c_d\! \cdot\! \bigg(\! \text{tanh}\!\bigg[ 2c_d \frac{\Real{\bar{d}_{n,k}^{(i)}}}{\bar{\sigma}_{d:{n,k}}^{2(i)}} \bigg]\!\! +\! j\text{tanh}\!\bigg[ 2c_d \frac{\Imag{\bar{d}_{n,k}^{(i)}}}{\bar{\sigma}_{{d}:{n,k}}^{2(i)}} \bigg]\!\bigg),\!\!
\label{eq:QPSK_denoiser_P1}
\end{equation}
where $c_d \triangleq \sqrt{E_\mathrm{D}/2}$ denotes the magnitude of the real and imaginary parts of the explicitly chosen \ac{QPSK} symbols, with its corresponding variance updated as in equation \eqref{eq:MSE_d_k_P1}.


After obtaining $\hat{d}_{n,k}^{(i)}$ as per eq. \eqref{eq:QPSK_denoiser_P1}, the final outputs are computed by damping the results with damping factors $0 < \beta_d < 1$ in order to improve convergence \cite{Su_TSP_2015}, yielding
\begin{equation}
\label{eq:d_damped_P1}
\hat{d}_{n,k}^{(i)} = \beta_d \hat{d}_{n,k}^{(i)} + (1 - \beta_d) \hat{d}_{n,k}^{(i-1)}.
\end{equation}

In turn, the corresponding variances $\hat{\sigma}^{2(i)}_{d:{n,k}}$ are first correspondingly updated via eq. \eqref{eq:MSE_d_k_P1} and then damped via
\begin{equation}
\label{eq:MSE_d_m_damped_P1}
\hat{\sigma}^{2(i)}_{d:{n,k}} = \beta_d \hat{\sigma}_{d:{n,k}}^{2(i)} + (1-\beta_d) \hat{\sigma}_{d:{n,k}}^{2(i-1)}.
\end{equation}

Finally, the consensus update can be obtained as
\begin{equation}
\label{eq:d_hat_final_est_P1}
\hat{d}_{k} = \frac{1}{N} \sum_{n=1}^N \hat{d}_{n,k}^{(i_\text{max})}.
\end{equation}

\subsection{Linear GaBP Derivation for Combiner Design}
\label{subsec:GaBP_combining}

For convenience, let us start by restating the combiner in eq. \eqref{eq:u_def_precoder} as
\vspace{-1ex}
\begin{equation}
\label{eq:u_def_precoder_v2_implicit}
\vspace{-1ex}
\bm{u} \!=\! (\underbrace{\bm{H}(\overbrace{\sigma_s^2\mathbf{I}_{K} + \boldsymbol{\Omega}}^{\bm{C} \;\in\; \mathbb{C}^{K \times K}} )\bm{H}\herm+\sigma^2_w \mathbf{I}_{N}}_{\bm{A} \;\in\; \mathbb{C}^{N \times N}})^{-1} \overbrace{\bm{H} \sigma_s^2 \mathbf{1}_{K}}^{\bm{b} \;\in\; \mathbb{C}^{N \times 1}} \!=\! \bm{A}^{-1} \bm{b},
\end{equation}
where we have implicitly defined the quantities $\bm{A} \;\in\; \mathbb{C}^{N \times N}$ and $\bm{b} \;\in\; \mathbb{C}^{N \times 1}$.

As an intermediate complexity reduction, let us apply the matrix inversion lemma \cite{RanasingheARXIV2024}, such that the equation \eqref{eq:u_def_precoder_v2_implicit} can be reexpressed as
\vspace{-1ex}
\begin{equation}
\vspace{-1ex}
\label{eq:u_def_precoder_v2_implicit_Woodbury}
\bm{u} = \frac{1}{\sigma^2_w}\big(\mathbf{I}_N - \bm{H}\big(\sigma^2_w\bm{C}^{-1} + \bm{H}\herm \bm{H}\big)^{-1} \bm{H}\herm \big) \bm{b}.
\end{equation}

Notice that this operation has now effectively reduced the computational complexity from $\mathcal{O}(N^3)$ to $\mathcal{O}(NK^2)$.

However, while this is quite useful in underloaded systems where $K < N$, this no longer demonstrates an advantage in fully-loaded systems when $K \approx N$, motivating us to consider a message passing based estimation to completely remove the matrix inversion.
Let us now consider the linear form extracted from equation \eqref{eq:u_def_precoder_v2_implicit} as
\vspace{-1ex}
\begin{equation}
\vspace{-1ex}
\label{eq:u_def_precoder_linear_form}
\bm{b} = \bm{A} \bm{u},
\end{equation}
where the goal is to estimate the combining vector, hereafter denoted $\hat{\bm{u}}$, given $\bm{A}$ and $\bm{b}$.

As one can see, this is identical to the linear form used in equation \eqref{eq:received_signal_matrix_ICC_cov_noise}, albeit without the noise term, which implies that the same \ac{GaBP} algorithm can be used to estimate $\bm{u}$.

Before a detailed derivation of the \ac{GaBP} rules, let us consider some of the statistical properties of the elements of the vector $\bm{u}$ to be estimated. 
The expectation can easily be extracted as $\mu_u = \mathbb{E}[\bm{u}] \approx 0$ due to the zero-mean Gaussian construction and \ac{iid} assumptions, while the variance of $\bm{u}$ is usually constrained and hence, $\sigma_u^2 \approx 1$.
These properties can now be leveraged to initialize the \ac{GaBP} algorithm, with the estimate for $\mu_u$, hereafter denoted as $\hat{\mu}_u^{(i)}$, updated iteratively according to the \ac{EM} algorithm detailed in Section \ref{subsubsec:EM_for_combiner}.

Element-wise, equation \eqref{eq:u_def_precoder_linear_form} can be rewritten as
\begin{equation}
\label{eq:u_def_precoder_linear_form_elementwise}
b_n = \sum_{n'=1}^N a_{n,n'} u_{n'}.
\end{equation}

Similarly to the latter subsection, given an $i$-th iteration of the algorithm, we denote the soft replicas of the $n'$-th combining element with the $n$-th modified channel vector $b_n$ at the previous iteration by $\hat{u}_{n,n'}^{(i-1)}$.

Then, the corresponding \acp{MSE} of these estimates for the $i$-th iteration are given by
\begin{equation}
\hat{\sigma}^{2(i)}_{u:{n,n'}} \triangleq \mathbb{E}_{\mathrm{u}_{n,n'}} \big[ | u_{n'} - \hat{u}_{n,n'}^{(i-1)} |^2 \big],
\label{eq:MSE_u_k_P1}
\end{equation}
$\forall (n,n')$, where $\mathbb{E}_{\mathrm{u}_{n,n'}}$ refers to the expectation over all the realizations.

\subsubsection{Soft Interference Cancellation} Exploiting equation \eqref{eq:u_def_precoder_linear_form_elementwise}, the \ac{sIC} symbols for the combining signals are given by
\begin{equation}
\label{eq:u_soft_IC_P1}
\begin{aligned}
\tilde{b}_{u:n,n'}^{(i)} &= b_{n} - \sum_{q \neq n'} a_{n,q}\hat{u}_{n,q}^{(i-1)} , \\
&= a_{n,n'} u_{n'} + \underbrace{\sum_{q \neq n'} a_{n,q}(u_q - \hat{u}_{n,q}^{(i-1)})}_\text{interference term}. 
\end{aligned}
\end{equation}

In turn, leveraging \ac{SGA} to approximate the interference and noise terms as Gaussian noise, the conditional \ac{PDF} of the \ac{sIC} symbols are given by
\begin{equation}
\label{eq:cond_PDF_u_P1}
\mathbb{P}_{\tilde{\mathrm{b}}_{\mathrm{u}:n,n'}^{(i)} \mid \mathrm{u}_{n'}}(\tilde{b}_{u:n,n'}^{(i)}|u_{n'}) \propto \mathrm{exp}\bigg[ -\frac{|\tilde{b}_{u:n,n'}^{(i)} - a_{n,n'} u_{n'}|^2}{\tilde{\sigma}_{u:n,n'}^{2(i)}} \bigg],
\end{equation}

\begin{algorithm}[H]
\caption{Joint Data Detection \& \ac{OTAC} for Single-Stream Integrated  Communication and Computing Systems}
\label{alg:proposed_ICC_new_P1}
\setlength{\baselineskip}{11pt}
\textbf{Input:} receive signal vector $\bm{y}\in\mathbb{C}^{N\times 1}$, complex channel matrix $\bm{H}\in\mathbb{C}^{N\times K}$, maximum number of iterations $i_{\max}$, data constellation power $E_\mathrm{D}$, noise variance $\sigma^2_w$ and damping factor $\beta_d, \beta_u$. \\
\textbf{Output:} $\hat{\bm{d}}$ and $\hat{f}(\bm{s})$ 
\vspace{-2ex} 
\begin{algorithmic}[1]  
\STATEx \hspace{-3.5ex}\hrulefill
\STATEx \hspace{-3.5ex}\textbf{Initialization}
\STATEx \hspace{-3.5ex} - Set iteration counter to $i=0$ and amplitudes $c_d = \sqrt{E_\mathrm{D}/2}$.
\STATEx \hspace{-3.5ex} - Set initial data estimates to $\hat{d}_{n,k}^{(0)} = 0$ and corresponding 
\STATEx \hspace{-2ex} variances to $\hat{\sigma}^{2(0)}_{d:{n,k}} = E_\mathrm{D}, \forall n,k$.
\STATEx \hspace{-3.5ex} - Set $\sigma_s^2 = \frac{E_\mathrm{D}}{K}$, $\sigma^2_u = 1$ and $\mu_u^{(i)} = 0$.
\STATEx \hspace{-3.5ex}\hrulefill
\STATEx \hspace{-3.5ex}\textbf{GaBP (Communication) Stage}: $\forall n, k$
\STATEx \hspace{-3.5ex}\textbf{for} $i=1$ to $i_\text{max}$ \textbf{do}
\STATE Compute \ac{sIC} data signal $\tilde{y}_{d:{n,k}}^{(i)}$ and its corresponding variance $\tilde{\sigma}^{2(i)}_{d:{n,k}}$ from equations \eqref{eq:d_soft_IC_P1} and \eqref{eq:soft_IC_var_d_P1}.
\STATE Compute extrinsic data signal belief $\bar{d}_{n,k}^{(i)}$ and its corresponding variance $\bar{\sigma}_{d:{n,k}}^{2(i)}$ from equations \eqref{eq:extrinsic_mean_d_P1} and \eqref{eq:extrinsic_var_d_P1}.
\STATE Compute denoised and damped data signal estimate $\hat{d}_{n,k}^{(i)}$ from equations \eqref{eq:QPSK_denoiser_P1} and \eqref{eq:d_damped_P1}.
\STATE Compute denoised and damped data signal variance $\hat{\sigma}_{d:{n,k}}^{2(i)}$ from equations \eqref{eq:MSE_d_k_P1} and \eqref{eq:MSE_d_m_damped_P1}.

\STATEx \hspace{-3.5ex}\textbf{end for}
\STATE Calculate $\hat{d}_k, \forall k$ (equivalently $\hat{\bm{d}}$) using equation \eqref{eq:d_hat_final_est_P1}.
\STATE Compute $\bm{A}$ and $\bm{b}$ from equation \eqref{eq:u_def_precoder_v2_implicit}.
\STATEx \hspace{-3.5ex}\textbf{GaBP (Combining) Stage}: $\forall n, n'$
\STATEx \hspace{-3.5ex}\textbf{for} $i=1$ to $i_\text{max}$ \textbf{do}
\STATE Compute \ac{sIC} combining signal $\tilde{b}_{u:{n,n'}}^{(i)}$ and its corresponding variance $\tilde{\sigma}^{2(i)}_{u:{n,n'}}$ from equations \eqref{eq:u_soft_IC_P1} and \eqref{eq:soft_IC_var_u_P1}.
\STATE Compute extrinsic beliefs $\bar{u}_{n,n'}^{(i)}$ and its corresponding variance $\bar{\sigma}_{u:{n,n'}}^{2(i)}$ from equations \eqref{eq:extrinsic_mean_u_P1} and \eqref{eq:extrinsic_var_u_P1}.
\STATE Compute denoised and damped combiner estimate $\hat{u}_{n,n'}^{(i)}$ from equations \eqref{eq:u_denoiser_mean_multi} and \eqref{eq:u_damped}.
\STATE Compute denoised and damped combiner variance $\hat{\sigma}_{u:{n,n'}}^{2(i)}$ from equations \eqref{eq:u_denoiser_var_mullti} and \eqref{eq:MSE_u_m_damped}.
\STATE Update $\hat{\mu}_u^{(i)}$ using equation \eqref{eq:EM_update_u}.

\STATEx \hspace{-3.5ex}\textbf{end for}
\STATE Compute $\hat{u}_{n'}, \forall n'$ using equation \eqref{eq:u_hat_final_est}.
\STATE Compute $\hat{f}(\bm{s})$ from equation \eqref{eq:Aircomp_SIC}.

\end{algorithmic}
\end{algorithm}
\vspace{-2ex}

\noindent with its conditional variances expressed as
\begin{equation}
\label{eq:soft_IC_var_u_P1}
\tilde{\sigma}_{u:n,n'}^{2(i)} = \sum_{q \neq n'} \left|a_{n,q}\right|^2 \hat{\sigma}^{2(i)}_{u:{n,q}}.
\end{equation}

\subsubsection{Belief Generation} Similarly, as a consequence of \ac{SGA} and in hand of the conditional \acp{PDF}, the extrinsic \acp{PDF} are obtained as
\vspace{-1ex}
\begin{equation}
\label{eq:extrinsic_PDF_u_P1}
\vspace{-1ex}
\prod_{q \neq n} \mathbb{P}_{\tilde{\mathrm{b}}_{\mathrm{u}:q,n'}^{(i)} \mid \mathrm{u}_{n'}}(\tilde{b}_{u:q,n'}^{(i)}|u_{n'}) \propto \mathrm{exp}\bigg[ - \frac{(u_{n'} - \bar{u}_{n,n'}^{(i)})^2}{\bar{\sigma}_{u:n,n'}^{2(i)}} \bigg],
\end{equation}
where the corresponding extrinsic means are defined as
\vspace{-1ex}
\begin{equation}
\label{eq:extrinsic_mean_u_P1}
\vspace{-1ex}
\bar{u}_{n,n'}^{(i)} = \bar{\sigma}_{u:n,n'}^{(i)} \sum_{q \neq n} \frac{a^*_{q,n'} \cdot \tilde{b}_{u:q,n'}^{(i)}}{ \tilde{\sigma}_{u:q,n'}^{2(i)}},
\end{equation}
with the extrinsic variances given by
\begin{equation}
\label{eq:extrinsic_var_u_P1}
\bar{\sigma}_{u:n,n'}^{2(i)} = \bigg( \sum_{q \neq n} \frac{|a_{q,n'}|^2}{\tilde{\sigma}_{u:q,n'}^{2(i)}} \bigg)^{\!\!\!-1}.
\end{equation}

\subsubsection{Soft Replica Generation} Since the combining signal elements follow a Gaussian distribution, the denoiser with a Gaussian prior and its corresponding variance is given by 
\begin{subequations}
\begin{equation}
\label{eq:u_denoiser_mean_multi}
\hat{u}_{n,n'}^{(i)} = \frac{\sigma^2_u \cdot \bar{u}_{n,n'}^{(i)} + \bar{\sigma}_{u:n,n'}^{2(i)} \cdot \hat{\mu}_u^{(i)}}{\bar{\sigma}_{u:n,n'}^{2(i)} + \sigma^2_u},
\end{equation}
\begin{equation}
\label{eq:u_denoiser_var_mullti}
\hat{\sigma}_{u:n,n'}^{2(i)} = \frac{\sigma^2_u \cdot \bar{\sigma}_{u:n,n'}^{2(i)}}{\bar{\sigma}_{u:n,n'}^{2(i)} + \sigma^2_u},
\end{equation}
\end{subequations}
where we reiterate that the estimate $\hat{\mu}_u^{(i)}$ can be updated via the \ac{EM} algorithm as detailed in the latter Section \ref{subsubsec:EM_for_combiner}.

After the denoiser, the estimates can be damped as
\begin{subequations}
\begin{equation}
\label{eq:u_damped}
\hat{u}_{n,n'}^{(i)} = \beta_u \hat{u}_{n,n'}^{(i)} + (1 - \beta_u) \hat{u}_{n,n'}^{(i-1)},
\end{equation}
\begin{equation}
\label{eq:MSE_u_m_damped}
\hat{\sigma}^{2(i)}_{u:{n,n'}} = \beta_u \hat{\sigma}_{u:{n,n'}}^{2(i)} + (1-\beta_u) \hat{\sigma}_{u:{n,n'}}^{2(i-1)}.
\end{equation}
\end{subequations}

The final consensus update can be obtained as
\begin{equation}
\label{eq:u_hat_final_est}
\hat{u}_{n'} =  \bigg( \sum_{n=1}^N \frac{|a_{n,n'}|^2}{\tilde{\sigma}_{u:n,n'}^{2(i)}} \bigg)^{\!\!\!-1} \! \! \bigg( \sum_{n=1}^N \frac{a^*_{n,n'} \cdot \tilde{b}_{u:n,n'}^{(i)}}{ \tilde{\sigma}_{u:n,n'}^{2(i)}} \bigg).
\end{equation}

\subsubsection{Expectation Maximization Update}
\label{subsubsec:EM_for_combiner}

Similarly to Section \ref{subsubsec:EM_alg}, under the assumption of Gaussian-distributed variables, the \ac{EM} update rule can be expressed as
\begin{equation}
\label{eq:EM_update_u}
\hat{\mu}_u^{(i)} = \frac{1}{N^2} \sum_{n=1}^N \sum_{n'=1}^N \hat{u}_{n,n'}^{(i)}.
\end{equation}

\vspace{-2ex}
\subsection{Performance and Complexity Analysis}


In this subsection, the performance and complexity analysis of the proposed single-stream \ac{ICC} framework are briefly evaluated.
For the numerical simulations, we keep the same system parameters as specified in Section \ref{subsec:performance_analysis_benchmarking}.

\begin{figure}[H]
\subfigure[{\footnotesize $N=100$}]%
{\includegraphics[width=0.9925\columnwidth]{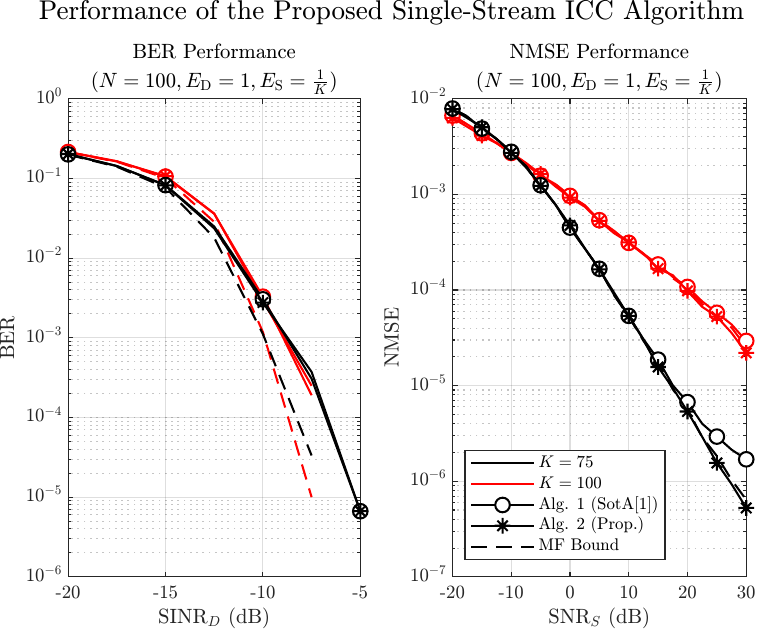}
\label{fig:BER_MSE_plot_1_ALg2}}\vspace{-1ex}\\
\vspace{-1ex}
\subfigure[{\footnotesize $N=200$}]%
{\includegraphics[width=0.9925\columnwidth]{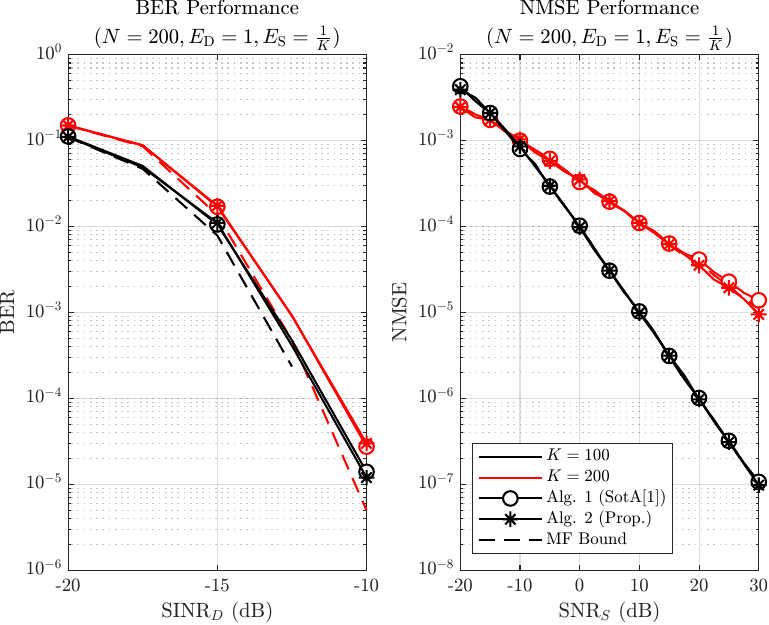}
\label{fig:BER_MSE_plot_2_ALg2}}
\vspace{-2ex}
\caption{\ac{BER} and \ac{NMSE} achieved by Algorithm \ref{alg:proposed_ICC_new_P1} in underloaded and fully-loaded scenarios, under two different system sizes.}
\label{fig:BER_MSE_Alg2}
\vspace{1ex}
\includegraphics[width=0.9925\columnwidth]{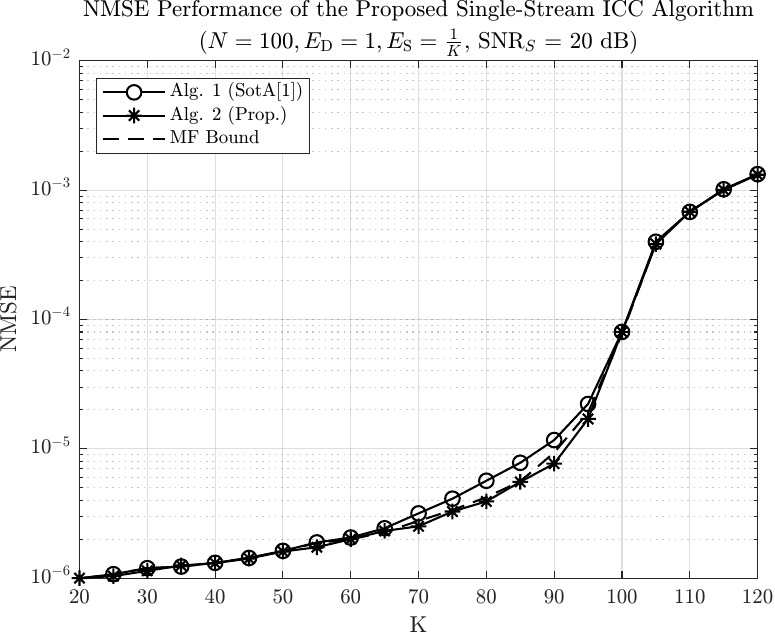}
\vspace{-4ex}
\caption{\ac{NMSE} achieved by Algorithms \ref{alg:proposed_ICC} and \ref{alg:proposed_ICC_new_P1}, as a function of the user load ($20\leq K \leq 120$), in a system of size $N=100$.}
\label{fig:MSE_plot_Alg2_vary_K}
\vspace{-2ex}
\end{figure}

It can be seen from Fig. \ref{fig:BER_MSE_Alg2} that both the \ac{BER} and \ac{NMSE} performances of the proposed single-stream scheme summarized in Algorithm \ref{alg:proposed_ICC_new_P1} are fundamentally identical to that of the benchmarking Algorithm \ref{alg:proposed_ICC}.
This is in spite of the fact that in this scheme no cancellation of the computing stream from the data detection is performed.

However, unlike in the previous benchmarking scheme with its prohibitive matrix inversions, the proposed scheme of Algorithm \ref{alg:proposed_ICC_new_P1} is of order $\mathcal{O}(NK)$ for the communication \ac{GaBP} loop and of order $\mathcal{O}(N^2)$ for the combining \ac{GaBP} loop.

Finally, it is worth noting from the results of Figures \ref{fig:BER_MSE_Alg1} and \ref{fig:BER_MSE_Alg2}, that indeed the performance of the \ac{OTAC} operation under the \ac{GaBP}-based framework for \ac{ICC} here considered indeed degrades as the loading conditions approach and surpass the fully-load scenario, which as explained earlier, can be attributed to the combiner described in equation \eqref{eq:u_def_precoder}.

In order to provide an empirical evidence to the latter, we offer in Figure \ref{fig:MSE_plot_Alg2_vary_K} a plot showing the \ac{NMSE} performances of Algorithms \ref{alg:proposed_ICC} and \ref{alg:proposed_ICC_new_P1}, as well as the corresponding \ac{MF} bound, as a function of the load $K$ in a system with $N=100$ antennas at the \ac{BS}/\ac{AP}.
It is confirmed that the proposed \ac{GaBP} is indeed very effective in allowing \ac{OTAC} in an integrated fashion, as long as the load in terms of data streams is not too high.

The results also motivate, however, the question as to whether the \ac{GaBP} detection framework for \ac{ICC} supports overloading in terms of the combined number of data and computing streams.
In order to frontally address this question, a generalization of the technique to a multi-stream \ac{OTAC} scenario is required, which is the aim of the next section.

\vspace{-0.5ex}
\section{Proposed Multi-Stream/Access ICC Framework}
\label{sec:multi_stream_ICC}

\subsection{Multi-Stream System Model}
\label{subsec:multi_stream_system_model}

In this section, we extend the \ac{ICC} framework described above to a scenario where the \ac{BS}/\ac{AP} aims to compute a vector of $M$ distinct and uncorrelated \ac{OTAC} target  functions $\bm{f}({\bm{s}}) \triangleq [f_1(\bm{s}_1), f_2(\bm{s}_2), \cdots, f_M(\bm{s}_M)]\trans \in \mathbb{C}^{M \times 1}$.

It will be assumed that the target functions are uncorrelated\footnote{Relaxing the uncorrelated assumption is possible, but requires the derivation of dedicated message-passing rules based on conditional probabilities that account for the correlation, and therefore is beyond the scope of this article.}, which imply that each computing signal only contributes to one target function.
It will also be assumed that all $K$ \acp{UE}/\acp{ED} contribute to one and only one \ac{OTAC} operation\footnote{This assumption is adopted to the disadvantage of the proposed scheme, as it implies a harder loading condition then the alternative where each \acp{UE}/\acp{ED} may or may not contribute to an \ac{OTAC} operation. Other than that, the assumption has no implication onto the proposed method to be described hereafter.}.

To further clarify the model, since the design of orthogonal precoders to separate each stream is not feasible due to the limited number on transmit antennas in a \ac{SIMO} system, we consider the computation of distinct target functions from the same set of computing symbols with a larger power allocation to the computing symbols for each function to be computed.

\begin{figure}[H]
\centering
\includegraphics[width=0.9\columnwidth]{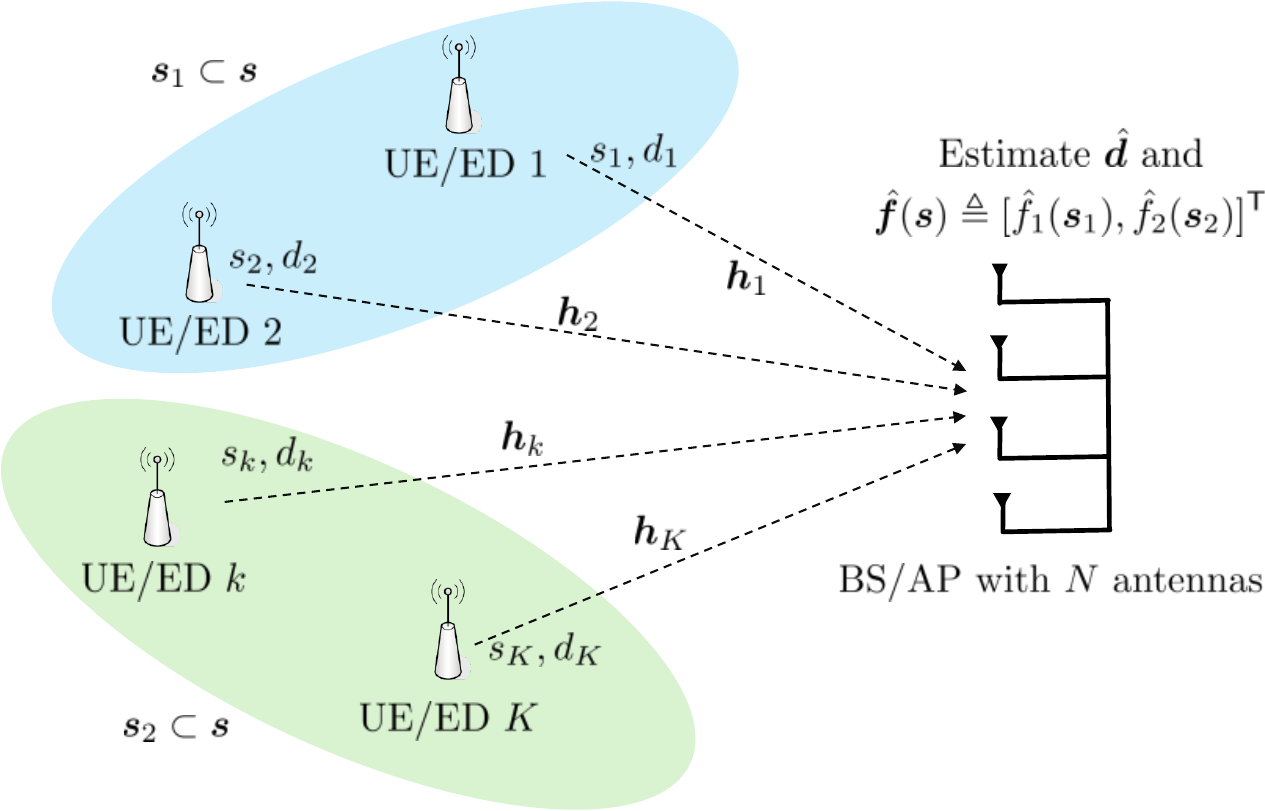}
\vspace{-1ex}
\caption{\ac{SIMO} \ac{ICC} system computing a vector of two distinct functions ($i.e., M = 2$) from the same set of computing symbols.}
\label{fig:system_model_multi-stream}
\vspace{-2ex}
\end{figure}

For the sake of clarity of exposition, let us consider an example as illustrated in Fig. \ref{fig:system_model_multi-stream} where the \ac{BS}/\ac{AP} aims to compute estimates for each of the two distinct target functions
\begin{equation}
\label{eq:target_function_multi_1}
f_1(\bm{s}_1)= \sum_{k=1}^{k'} s_k, \;\; \text{and} \;\; f_2(\bm{s}_2) = \sum_{k=k' + 1}^{K} s_k,
\end{equation}
where $k' \triangleq \lfloor K/2 \rfloor$ and $\lfloor \cdot \rfloor$ denotes the floor operation.

Referring to equation \eqref{eq:Aircomp_SIC}, and generalizing it to the multi-stream case we have
\begin{equation}
\vspace{-1ex}
\label{eq:Aircomp_SIC_MS}
\hat{f}_m(\bm{s}_m) = \bm{u}_m\herm(\bm{y} - \bm{H}\hat{\bm{d}}) = \bm{u}_m\herm(\bm{H} (\bm{s} - \check{\bm{d}}) +\bm{w}),
\end{equation}
where the combining vectors $\bm{u}_m$ can be obtained by leveraging a similar formulation as in equation \eqref{eq:min_problem} into $m$ separate minimization problems, namely
\begin{equation}
\label{eq:min_problem_1}
\underset{\bm{u}_m\in\mathbb{C}^{N\times1}}{\mathrm{minimize}} \hspace{3ex} \mathbb{E} \big[ \| \bm{p}_{m}\trans \bm{s} \!-\! \bm{u}_m\herm(\bm{H} (\bm{s} \!-\! \check{\bm{d}}) +\bm{w})  \|_2^2 \big],
\end{equation}
where $\bm{p}_m$ is a sparse vector containing ones in the positions of \acp{UE}/\acp{ED} that contribute to the $m$-th \ac{OTAC} target function.

For instance, in the example given in equation \eqref{eq:target_function_multi_1},
\begin{equation}
\label{eq:p1_p2_def}
\bm{p}_1 = [1, \dots, 1, 0, \dots, 0]\trans \in \mathbb{C}^{K \times 1},
\end{equation}
and
\begin{equation}
\label{eq:p1_p2_def_2}
\bm{p}_2 = [0, \dots, 0, 1, \dots, 1]\trans \in \mathbb{C}^{K \times 1},
\end{equation}
with the first $k'$-th and last $K-k'$-th elements set to one, respectively.

The corresponding combining vectors $\bm{u}_m$ are given by
\vspace{-1ex}
\begin{equation}
\label{eq:u_def_precoder_per_n_multi}
\vspace{-1ex}
\bm{u}_m = (\underbrace{\bm{H}(\sigma_s^2\mathbf{I}_{K} + \boldsymbol{\Omega} )\bm{H}\herm+\sigma^2_w \mathbf{I}_{N}}_{\bm{A} \;\in\; \mathbb{C}^{N \times N}})^{-1} \underbrace{\bm{H} \sigma_s^2 \bm{p}_m}_{\bm{b}_m \;\in\; \mathbb{C}^{ \times 1}},
\end{equation}
which in turn implies that
\vspace{-1ex}
\begin{equation}
\label{eq:u_def_precoder_linear_form}
\bm{b}_m = \bm{A} \bm{u}_m,
\end{equation}
such that the same \ac{GaBP} approach described in Subsection \ref{subsec:GaBP_combining} can be applied.

It is easy to see that the aforementioned framework can be extended to any number of distinct functions to be computed from the same set of computing symbols by simply defining the corresponding index vectors $\bm{p}_m$, and designing associated combiner vectors for each computing stream accordingly.
\vspace{-1ex}
\begin{algorithm}[H]
\caption{Joint Data Detection \& \ac{OTAC} for Multi-Stream Integrated  Communication and Computing Systems}
\label{alg:proposed_ICC_new_P2}
\setlength{\baselineskip}{11pt}
\textbf{Input:} receive signal vector $\bm{y}\in\mathbb{C}^{N\times 1}$, complex channel matrix $\bm{H}\in\mathbb{C}^{N\times K}$, maximum number of iterations $i_{\max}$, data constellation power $E_\mathrm{D}$, noise variance $\sigma^2_w$ and damping factor $\beta_d, \beta_u$. \\
\textbf{Output:} $\hat{\bm{d}}$ and $\hat{\bm{f}}(\bm{s})$.
\vspace{-2ex} 
\begin{algorithmic}[1]  
\STATEx \hspace{-3.5ex}\hrulefill
\STATEx \hspace{-3.5ex}\textbf{Initialization}
\STATEx \hspace{-3.5ex} - Set iteration counter to $i=0$ and amplitudes $c_d = \sqrt{E_\mathrm{D}/2}$.
\STATEx \hspace{-3.5ex} - Set initial data estimates to $\hat{d}_{n,k}^{(0)} = 0$ and corresponding 
\STATEx \hspace{-2ex} variances to $\hat{\sigma}^{2(0)}_{d:{n,k}} = E_\mathrm{D}, \forall n,k$.
\STATEx \hspace{-3.5ex} - Set $\sigma_s^2 = \frac{E_\mathrm{D}}{K}$, $\sigma^2_u = 1$ and $\mu_u^{(i)} = 0$.
\STATEx \hspace{-3.5ex}\hrulefill
\STATEx \hspace{-3.5ex}\textbf{Linear GaBP (Communication) Stage}: $\forall n, k$
\STATEx \hspace{-3.5ex}\textbf{for} $i=1$ to $i_\text{max}$ \textbf{do}
\STATE Compute \ac{sIC} data signal $\tilde{y}_{d:{n,k}}^{(i)}$ and its corresponding variance $\tilde{\sigma}^{2(i)}_{d:{n,k}}$ from equations \eqref{eq:d_soft_IC_P1} and \eqref{eq:soft_IC_var_d_P1}.
\STATE Compute extrinsic data signal belief $\bar{d}_{n,k}^{(i)}$ and its corresponding variance $\bar{\sigma}_{d:{n,k}}^{2(i)}$ from equations \eqref{eq:extrinsic_mean_d_P1} and \eqref{eq:extrinsic_var_d_P1}.
\STATE Compute denoised and damped data signal estimate $\hat{d}_{n,k}^{(i)}$ from equations \eqref{eq:QPSK_denoiser_P1} and \eqref{eq:d_damped_P1}.
\STATE Compute denoised and damped data signal variance $\hat{\sigma}_{d:{n,k}}^{2(i)}$ from equations \eqref{eq:MSE_d_k_P1} and \eqref{eq:MSE_d_m_damped_P1}.

\STATEx \hspace{-3.5ex}\textbf{end for}
\STATE Calculate $\hat{d}_k, \forall k$ (equivalently $\hat{\bm{d}}$) using equation \eqref{eq:d_hat_final_est_P1}.
\STATE Compute $\bm{A}$ from equation \eqref{eq:u_def_precoder_per_n_multi}.
\STATEx \hspace{-3.5ex}\textbf{GaBP (Combining) Stage}: $\forall n, n', m$
\STATEx \hspace{-3.5ex}\textbf{for} $m=1$ to $M$ \textbf{do}
\STATE Compute $\bm{b}_m$ from equation \eqref{eq:u_def_precoder_per_n_multi}.
\STATEx \hspace{-2.5ex}\textbf{for} $i=1$ to $i_\text{max}$ \textbf{do}
\STATE Compute \ac{sIC} combining signal $\tilde{b}_{u:{n,n'}}^{(i)}$ and its corresponding variance $\tilde{\sigma}^{2(i)}_{u:{n,n'}}$ from equations \eqref{eq:u_soft_IC_P1} and \eqref{eq:soft_IC_var_u_P1}.
\STATE Compute extrinsic beliefs $\bar{u}_{n,n'}^{(i)}$ and its corresponding variance $\bar{\sigma}_{u:{n,n'}}^{2(i)}$ from equations \eqref{eq:extrinsic_mean_u_P1} and \eqref{eq:extrinsic_var_u_P1}.
\STATE Compute denoised and damped combiner estimate $\hat{u}_{n,n'}^{(i)}$ from equations \eqref{eq:u_denoiser_mean_multi} and \eqref{eq:u_damped}.
\STATE Compute denoised and damped combiner variance $\hat{\sigma}_{u:{n,n'}}^{2(i)}$ from equations \eqref{eq:u_denoiser_var_mullti} and \eqref{eq:MSE_u_m_damped}.
\STATE Update $\hat{\mu}_u^{(i)}$ using equation \eqref{eq:EM_update_u}.

\STATEx \hspace{-2.5ex}\textbf{end for}
\STATE Compute the $m$-th $\hat{u}_{n'}, \forall n'$ using equation \eqref{eq:u_hat_final_est}.
\STATE Calculate $\hat{f}_m(\bm{s}), \forall m$ using equation \eqref{eq:Aircomp_SIC_MS}.
\STATEx \hspace{-3.5ex}\textbf{end for}

\end{algorithmic}
\end{algorithm}
\vspace{-2ex}

As a consequence of the multi-stream combining framework defined in the previous subsection, a similar algorithmic approach can be leveraged to compute the combiner defined in equation \eqref{eq:u_def_precoder_per_n_multi}.
The resulting proposed procedure is summarized in Algorithm \ref{alg:proposed_ICC_new_P2}.

\vspace{-2ex}
\subsection{Performance Metrics}

Given the system model described in equation \eqref{eq:received_signal_matrix}, the previously defined single-stream \ac{SINR}/\ac{SNR} definitions have to be extended to the multi-stream case.
In particular, we have
\begin{equation}
\label{eq:SINR_def_multi}
\text{SINR}_D^M \triangleq \frac{\mathbb{E}[|| \bm{H} \bm{d} ||^2]}{M \mathbb{E}[|| \bm{H} \bm{s} ||^2] + \sigma^2_w},
\end{equation}
and
\begin{equation}
\label{eq:SINR_S_def_multi}
\text{SNR}_S^M \!\triangleq\! \frac{M \mathbb{E}[|| \bm{H} \bm{s} ||^2]}{\sigma^2_w},
\end{equation}
where $M$ is the number of distinct functions being computed.

\vspace{-2ex}
\subsection{Performance and Complexity Analysis}

We now analyze the performance of the proposed multi-stream \ac{ICC} framework proposed in Algorithm \ref{alg:proposed_ICC_new_P2}.
While the proposed algorithm is derived for an arbitrary $M$, we focus on the case where $M=2$ for the sake of simplicity and clarity, and also perform some comparisons with the previous cases to demonstrate the gain in performance.
\vspace{-2ex}
\begin{figure}[H]
\centering
\includegraphics[width=\columnwidth]{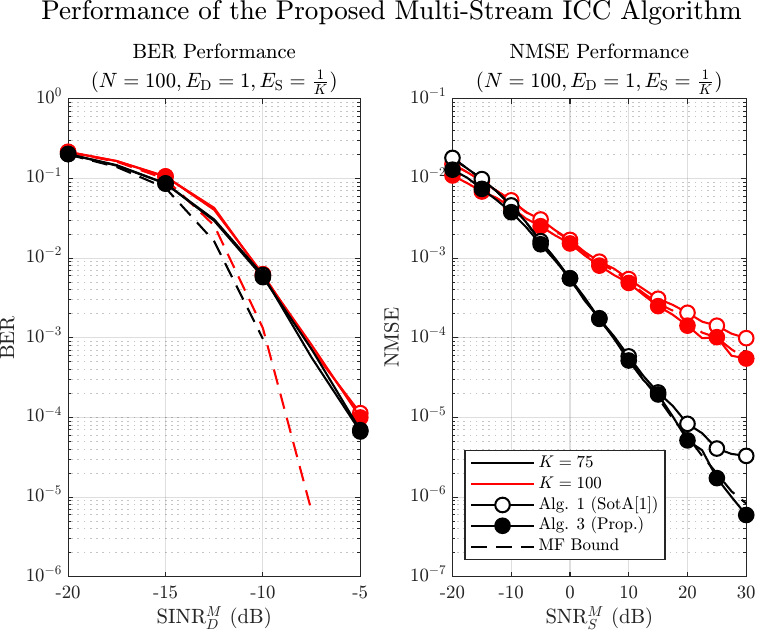}
\vspace{-4ex}
\caption{\ac{BER} and \ac{NMSE} performance of the proposed multi-stream Algorithm \ref{alg:proposed_ICC_new_P2} for the underloaded and fully-loaded scenarios in the multi-stream regime.}
\label{fig:BER_MSE_plot_Alg2_multistream}
\vspace{-2ex}
\end{figure}

As seen from Fig. \ref{fig:BER_MSE_plot_Alg2_multistream}, the legend for Algorithm \ref{alg:proposed_ICC} in the multi-stream regime represents the relevant algorithm utilized in conjunction with the new combiner proposed in Subsection \ref{subsec:GaBP_combining}.
The results show that the proposed multi-stream \ac{ICC} framework achieves the same performance in terms of both the \ac{BER} and \ac{NMSE} as the previously developed algorithms.

\subsection{Proposed Multi-Access ICC Framework}
\label{sec:multi_access_ICC}

Let $\mathcal{K}_D$, $\mathcal{K}_S$ and $\mathcal{K}_{DS}$ denote the sets of \ac{UE}/\ac{ED} indices corresponding to the communication, computing and \ac{ICC} operations, respectively, with $\mathcal{K}_D \cup \mathcal{K}_S \cup \mathcal{K}_{DS} = \{1, \dots, K\}$ and $|\mathcal{K}_D| + |\mathcal{K}_S| + |\mathcal{K}_{DS}| = K$, with the corresponding number of data, computing and \ac{ICC} users denoted by $K_D$, $K_S$ and $K_{DS}$, respectively.

\begin{figure}[H]
\centering
\includegraphics[width=0.9\columnwidth]{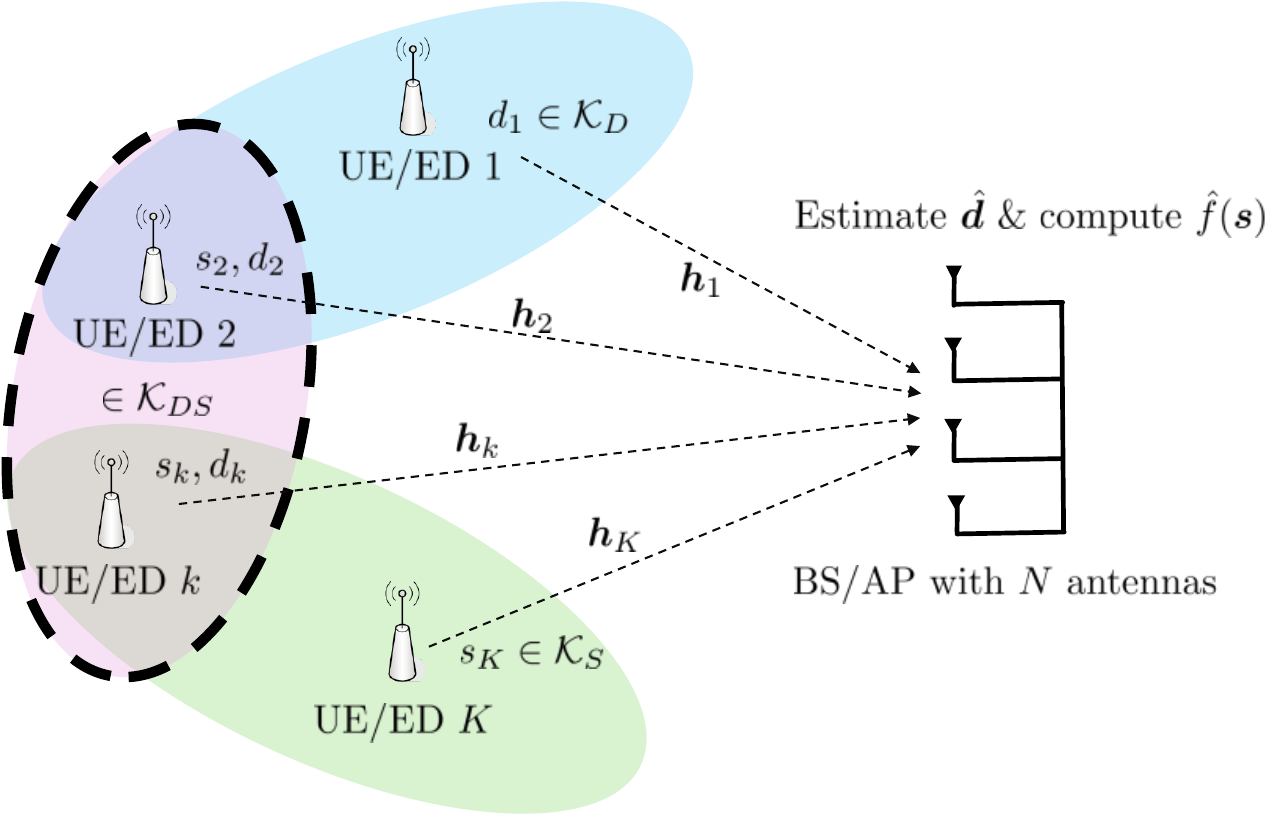}
\vspace{-1ex}
\caption{\ac{SIMO} \ac{ICC} system computing a vector of distinct functions from the same set of computing symbols where each \ac{UE}/\ac{ED} either transmits a data signal, computing signal or both.}
\label{fig:system_model_multi-access}
\vspace{-1ex}
\end{figure}

As an example, consider the case illustrated in Fig. \ref{fig:system_model_multi-access}, where $\mathcal{K}_D = \{ 1 \}$, $\mathcal{K}_S = \{ 2, \dots, k \}$ and $\mathcal{K}_{DS} = \{ k+1, \dots, K \}$. 
More concretely, if $K = 10$ and $k = 5$, then $\mathcal{K}_D = \{ 1 \}$, $\mathcal{K}_S = \{ 2, 3, 4, 5 \}$ and $\mathcal{K}_{DS} = \{ 6, 7, 8, 9, 10 \}$.

For enabling a multi-access receiver design, the key piece of additional information would be the knowledge of the set $\mathcal{K}_S$, which can be easily obtained via a pilot signal exchange or a pre-defined mapping.

In hand of this information and under the assumption that $\bm{s}$ is a low power signal, the only resulting change to all the proposed algorithms would be an enforcing condition on the data and data variance estimates inside the \ac{GaBP} framework; $i.e.,$ $\hat{d}_{n,k}^{(i)} = 0, \forall n,k \in \mathcal{K}_S, \mathcal{K}_{DS}$ and $\hat{\sigma}^{2(i)}_{d:{n,k}} = 0, \forall n,k \in \mathcal{K}_S, \mathcal{K}_{DS}$.

Let $K_C = K_S + K_{DS}$ denote the total number of computing users in the system and in line with the previous sections of the manuscript.

\begin{figure}[H]
\centering
\includegraphics[width=\columnwidth]{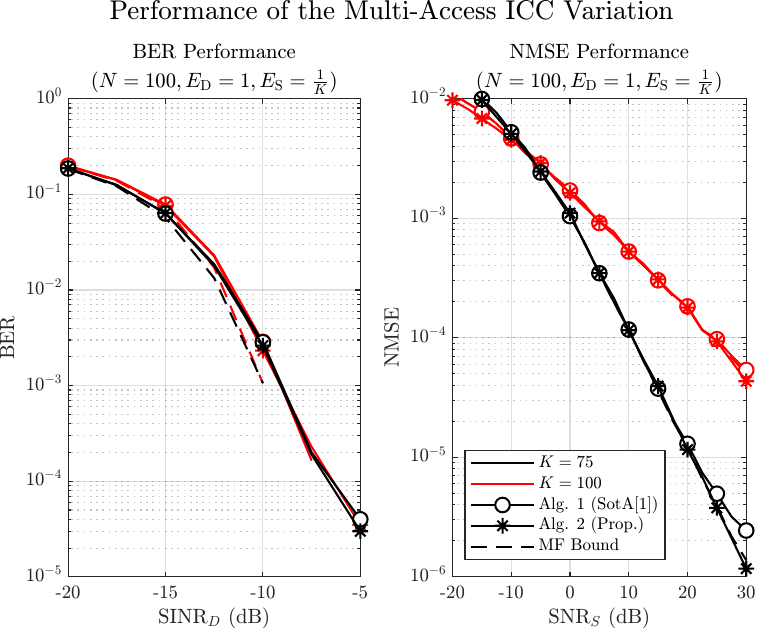}
\vspace{-3ex}
\caption{\ac{BER} and \ac{NMSE} performance of the proposed single-stream Algorithm \ref{alg:proposed_ICC_new_P1} for the underloaded and fully-loaded scenarios in the multi-access regime.}
\label{fig:BER_MSE_plot_Alg2_multiaccess}
\vspace{-2ex}
\end{figure}

Fig. \ref{fig:BER_MSE_plot_Alg2_multiaccess} shows the performance of the proposed multi-access \ac{ICC} framework in terms of \ac{BER} and \ac{NMSE} for the overloaded, underloaded and fully-loaded scenarios.
In the underloaded case, $K_D= 25$, $K_S = 25$ and $K_{DS} = 25$, in the fully-loaded case, $K_D= 30$, $K_S = 30$ and $K_{DS} = 40$, and in the overloaded case, $K_D= 40$, $K_S = 40$ and $K_{DS} = 45$. 

In terms of the results, we see that there's minimal performance degradation in all the loading scenarios even when the multi-access \ac{ICC} framework is utilized, which is a testament to the robustness of the proposed framework.

\section{Conclusion}
\label{sec:conclusion}

In this manuscript, we proposed several novel frameworks for the design of practical \ac{ICC} receivers, with an emphasis on the flexibility and scalability of the systems.
We first proposed a single-stream \ac{ICC} framework that leverages the \ac{GaBP} algorithm to detect the data signals and compute a single target function.
We then extended the framework to a multi-stream \ac{ICC} framework that computes multiple target functions from the same set of computing symbols.
Finally, we proposed a multi-access \ac{ICC} framework that enables the multiple users to either transmit communication symbols, computing signals or both with minimal change to the receiver structure.
The proposed frameworks were shown to reach the bounds set by the \ac{SotA} algorithm in terms of both \ac{BER} and \ac{NMSE} performance, while also being robust to the varying system loading scenarios with a low complexity.

\section{Appendix}
\label{sec:appendix}

\subsection{Derivation of the MMSE Estimator for equation \eqref{eq:min_problem}}
\label{sec:appendix_MMSE_main}

Let us start by noting that \ac{MMSE}-optimal vector $\bm{u} \in \mathbb{C}^{N \times 1}$ for the optimization problem in equation \eqref{eq:min_problem} can be recasted as
\begin{equation}
\label{eq:min_problem_exp}
\bm{u}_\text{MMSE} = \mathrm{arg} \; \underset{\bm{u}}{\mathrm{min}} \hspace{1ex} \mathbb{E} \big[ | \mathbf{1}_{K}\trans \bm{s} \!-\! \bm{u}\herm\big(\bm{H} (\bm{s} \!-\! \check{\bm{d}}) +\bm{w}\big)  |^2 \big], 
\end{equation}
since this is a scalar optimization problem.

It is well known that the optimal solution to the \ac{MMSE} problem is given by
\begin{equation}
\label{eq:MMSE_solution}
\bm{u}_\text{MMSE} = \mathbf{R}^{-1} \mathbf{r},
\end{equation}
where 
\begin{equation}
\label{eq:cov_matrix}
\mathbf{R} = \mathbb{E} \big[ \big(\bm{H} (\bm{s} \!-\! \check{\bm{d}}) +\bm{w}\big) \big(\bm{H} (\bm{s} \!-\! \check{\bm{d}}) +\bm{w}\big)\herm \big],
\end{equation}
and
\begin{equation}
\label{eq:cross_cov_matrix}
\mathbf{r} = \mathbb{E} \big[ \big(\bm{H} (\bm{s} \!-\! \check{\bm{d}}) +\bm{w}\big) \bm{s}\herm \mathbf{1}_{K}  \big].
\end{equation}

Starting with the covariance matrix $\mathbf{R} \in \mathbb{C}^{N \times N}$ defined in equation \eqref{eq:cov_matrix}, we can write
\begin{align}
\label{eq:cov_matrix_exp}   
\mathbf{R} 
&= \mathbb{E} \big[ \big(\bm{H} (\bm{s} \!-\! \check{\bm{d}}) +\bm{w}\big) \big(\bm{H} (\bm{s} \!-\! \check{\bm{d}}) +\bm{w}\big)\herm \big] \nonumber \\
&= \mathbb{E} \big[ \bm{H} (\bm{s} \!-\! \check{\bm{d}}) (\bm{s} \!-\! \check{\bm{d}})\herm\bm{H}\herm + \bm{H} (\bm{s} \!-\! \check{\bm{d}}) \bm{w}\herm  \nonumber \\
&\hspace{5.5ex}+ \bm{w} (\bm{s} \!-\! \check{\bm{d}})\herm \bm{H}\herm + \bm{w}\bm{w}\herm  \big] \nonumber \\
&= \bm{H} \; \mathbb{E} \big[ \bm{s}\bm{s}\herm - \bm{s}\check{\bm{d}}\herm - \check{\bm{d}}\bm{s}\herm + \check{\bm{d}}\check{\bm{d}}\herm \big] \; \bm{H}\herm + \sigma^2_w \mathbf{I}_N \nonumber \\
&= \bm{H} (\sigma^2_s \mathbf{I}_K + \bm{\Omega}) \bm{H}\herm  + \sigma^2_w \mathbf{I}_N,
\end{align}
where the third equality follows from the zero-mean assumptions on the noise $\bm{w}$, computing signal $\bm{s}$ and channel matrix $\bm{H}$, and the last equality follows from the definition of the data error covariance matrix $\bm{\Omega} \triangleq \mathbb{E}[\check{\bm{d}}\check{\bm{d}}\herm]$.

Similarly, the cross-covariance vector $\mathbf{r} \in \mathbb{C}^{N \times 1}$ defined in equation \eqref{eq:cross_cov_matrix} can be expressed as
\begin{align}
\label{eq:cross_cov_matrix_exp}
\mathbf{r}
&= \mathbb{E} \big[ \big(\bm{H} (\bm{s} \!-\! \check{\bm{d}}) +\bm{w}\big) \bm{s}\herm \mathbf{1}_{K}  \big] \nonumber \\
&= \mathbb{E} \big[ \bm{H} \bm{s}\bm{s}\herm \mathbf{1}_{K}  - \bm{H} \check{\bm{d}} \bm{s}\herm \mathbf{1}_{K}  + \bm{w} \bm{s}\herm \mathbf{1}_{K}  \big] \nonumber \\
&= \bm{H} \sigma^2_s \mathbf{1}_{K}. 
\end{align}

Combining both terms defined in equations \eqref{eq:cov_matrix_exp} and \eqref{eq:cross_cov_matrix_exp} gives us the final expression
\begin{equation}
\label{eq:u_MMSE_expression}
\bm{u}_\text{MMSE} = \big(\bm{H} (\sigma^2_s \mathbf{I}_K + \bm{\Omega}) \bm{H}\herm  + \sigma^2_w \mathbf{I}_N\big)^{-1} \bm{H} \sigma^2_s \mathbf{1}_{K}.
\end{equation}

\subsection{Derivation of the MMSE Estimator for equation \eqref{eq:min_problem_1}}

Similarly to the previous section \ref{sec:appendix_MMSE_main}, we can recast the \ac{MMSE} optimization problem in equation \eqref{eq:min_problem_1} as
\begin{equation}
\label{eq:min_problem_exp_m}
\bm{u}_m = \mathrm{arg} \; \underset{\bm{u}_m}{\mathrm{min}} \hspace{1ex} \mathbb{E} \big[ | \bm{p}_m\trans \bm{s} \!-\! \bm{u}_m\herm\big(\bm{H} (\bm{s} \!-\! \check{\bm{d}}) +\bm{w}\big)  |^2 \big]. 
\end{equation}

As observed from equation \eqref{eq:min_problem_exp_m}, the only difference between the two optimization problems is the presence of the index vector $\bm{p}_m$ in the objective function instead of the vector $\mathbf{1}_{K}$ present in equation \eqref{eq:min_problem_exp}.

Therefore, we trivially obtain that the optimal solution to the \ac{MMSE} problem is given by
\begin{equation}
\label{eq:u_MMSE_expression_m}
\bm{u}_m = \big(\bm{H} (\sigma^2_s \mathbf{I}_K + \bm{\Omega}) \bm{H}\herm  + \sigma^2_w \mathbf{I}_N\big)^{-1} \bm{H} \sigma^2_s \bm{p}_m.
\end{equation}

\bibliographystyle{IEEEtran}
\bibliography{references}

\end{document}